\documentstyle[prd,aps,epsfig,twocolumn,floats,tabularx]{revtex}
\newcommand{\MM}{{\cal M}}
\newcommand{\NN}{{\cal N}}
\newcommand{\DD}{{\cal D}}
\newcommand{\SS}{{\cal S}}
\newcommand{\GG}{{\cal G}}
\newcommand{\cd}{\cdot}

\newcommand{\al}{\alpha}
\renewcommand{\b}{\beta}
\newcommand{\de}{\delta}
\newcommand{\De}{\Delta}
\newcommand{\ep}{\epsilon}
\newcommand{\ga}{\gamma}

\newcommand{\io}{\iota}
\newcommand{\La}{\Lambda}
\newcommand{\la}{\lambda}
\newcommand{\Om}{\Omega}
\newcommand{\om}{\omega}
\newcommand{\si}{\sigma}
\newcommand{\Si}{\Sigma}
\newcommand{\th}{\theta}
\newcommand{\Th}{\Theta}

\newcommand{\vph}{\varphi}
\newcommand{\ra}{\rightarrow}
\newcommand{\bk}{{\bf k}}
\newcommand{\bn}{{\bf n}}

\newcommand{\bm}[1]{\mbox{\boldmath $#1$}}

\newcommand{\be}{\begin{equation}}
\newcommand{\ee}{\end{equation}}
\newcommand{\bea}{\begin{eqnarray}}
\newcommand{\eea}{\end{eqnarray}}
\newcommand{\bean}{\begin{eqnarray*}}
\newcommand{\eean}{\end{eqnarray*}}
\newcommand{\dd}{\partial}

\newcommand{\gsim}{\stackrel{>}{\sim}}
\newcommand{\lsim}{\stackrel{<}{\sim}}
\newcommand{\Mpc}{{\rm Mpc}}
\renewcommand{\r}{\right}
\renewcommand{\l}{\left}
\title{Cosmic Microwave Background Anisotropies
 from Scaling Seeds: \\ Global Defect Models}
\author{R. Durrer, M. Kunz and A. Melchiorri}
\address{D\'epartement de Physique Th\'eorique,
         Universit\'e de Gen\`eve,
        24 quai Ernest Ansermet, CH-1211 Gen\`eve 4, Switzerland}
\begin{document}
\maketitle
\begin{abstract}\begin{center}
{\large\bf Abstract}
\end{center}
We investigate the global texture model of structure formation in
cosmogonies with non-zero cosmological constant for different
values of the Hubble parameter. We find that the absence of
significant acoustic peaks and little power on large scales
are robust predictions of these models. However, from a careful comparison
with data we conclude that at present we cannot safely reject the 
model on the grounds of present CMB data. Exclusion  by means of 
galaxy correlation data, requires assumptions on biasing and
statistics.  New, very stringent constraints come from peculiar
velocities.

Investigating the large-$N$ limit, we argue that our main conclusions
apply to all global $O(N)$ models of structure formation.
\end{abstract}

\section{Introduction}
Recently, a lot of effort has gone into the
determination of cosmological parameters from measurements of cosmic
microwave background (CMB) anisotropies, especially in view of the two
planned satellite experiments MAP and PLANCK\cite{PLMAP}. 
However, we believe it is important
to be aware of the heavy modeling which enters these results. In
general,  simple power law initial spectra for scalar
and tensor perturbations and vanishing vector perturbations are  assumed, as
predicted from inflation. To reproduce observational data,
the composition of the dark matter and the cosmological
parameters  as well as the input spectrum and the scalar to tensor
ratio are varied\cite{infla}.

We want to take a different approach: We modify the model for structure
formation. We assume that cosmic structure was induced by scaling
seeds. Using a simplified (and not very accurate) treatment for the photon
propagation, we have already shown that some key observations can be
reproduced within a very restricted family of scaling seed 
models\cite{observ}. Here we want to outline in detail a more accurate
computation with a fully gauge-invariant Boltzmann code 
especially adapted to treat models with sources. In this paper we
follow the philosophy of a general analysis of scaling seed models
motivated in Ref.~\cite{DK}. 

 Seeds are an
inhomogeneously distributed form of matter (like {\em e.g.} topological
defects) which interacts with the cosmic fluid only
gravitationally and which represents always a small fraction of the
total energy of the universe. They induce geometrical
perturbations, but their influence on the evolution of the background
universe can be neglected. Furthermore, in first order perturbation 
theory, seeds evolve according to the unperturbed spacetime geometry. 

Here, we mainly investigate the models of structure formation with global
texture. This models (for $\Om_{matter}=1$) show discrepancies with
the observed intermediate scale CMB anisotropies and with the galaxy
power spectrum on large scales\cite{PST}. Recently it has been argued 
that the addition of a
cosmological constant leads to better agreement with data for the 
cosmic string model of structure formation~\cite{Avel}. We analyze 
this question for the
texture model, by using ab initio simulation of cosmic texture as
described in Ref.~\cite{ZD}. We determine the CMB anisotropies, the
dark matter power spectrum and the bulk velocities for these models.
We also compare our results with the large-$N$
limit of global $O(N)$ models, and we discuss briefly which type of parameter
changes in the $2$-point functions of the seeds may lead to better 
agreement with data.

We find that the absence of significant acoustic peaks in the CMB
anisotropy spectrum is a robust result for global texture as well as
for the large-$N$ limit for all choices of cosmological parameters
investigated. Furthermore, the dark matter power spectrum on large
scales, $\la\gsim 20h^{-1}\Mpc$, is substantially lower than the
measured galaxy power spectrum.

However, comparing our CMB anisotropy spectra with present data,
we cannot safely reject the model.
On large angular scales, the CMB spectrum is in quite good
agreement with the COBE data set,  while on smaller scales 
we find a significant disagreement only with the Saskatoon
experiment. Furthermore, for non-satellite experiments foreground contamination
remains a serious problem due to the  limited sky and frequency coverage. 

The dark matter power spectra are clearly too low on large scales, but in
view of the unresolved biasing problem, we feel reluctant to rule out
the models on these grounds. A much clearer rejection may come from the
bulk velocity on large scales. Our prediction is by a factor 3 to 5 
lower than the POTENT result on large scales.

Since global texture and the large-$N$ limit lead to very similar
results, we conclude that all global $O(N)$ models of
structure formation for the cosmogonies investigated in this work
are ruled out if the bulk velocity on scales of $50h^{-1}$Mpc is
around 300km/s or if the CMB primordial anisotropies power spectrum
really shows a structure of peaks on sub-degree angular scales.

This paper is the first of a series of analyses of models with
scaling seeds. We therefore fully present the formalism used for our
calculations  in the next section. There, we also explain in detail
the  eigenvector expansion which allows to calculate the CMB
anisotropies and matter power spectra in models with seeds from the
two point functions of the seeds alone. This section can be
skipped if the reader is mainly interested in the results.
 Section~3 is devoted
to a brief description of the numerical simulations. In Section~4 we 
analyze our results and in Section~5 we draw some conclusions. 
 Two appendices are devoted to detailed definitions of the
perturbation variables and to some technical derivations.
\vspace{0.1cm}\\
{\bf Notation:}\hspace{3mm}
We always work in a spatially flat Friedmann universe. The metric is 
given by 
\[ds^2=a(t)^2(dt^2-\de_{ij}dx^idx^j)~,\]
 where $t$ denotes conformal time.

Greek indices denote spacetime coordinates $(0$ to $3)$ whereas Latin
ones run from 1 to 3. Three dimensional vectors are denoted by bold
face characters.

\section{The formalism}
Anisotropies in the CMB are small and can thus be described by first
order cosmological perturbation theory which we apply
throughout.  We neglect the non-linear evolution of density fluctuations 
 on smaller scales. Since models with seeds are genuinely non-Gaussian, the
usual numerical N-body simulations which start from Gaussian initial conditions
cannot be used to describe the evolution on smaller scales.

 Gauge-invariant perturbation equations for cosmological models
with seeds have been derived in Refs.~\cite{D90,Review}. 
Here we follow the notation and use the results presented in 
Ref.~\cite{Review}. Definitions of all the gauge-invariant perturbation
variables used here in terms of perturbations of the metric, the
energy momentum tensor and the brightness are given 
in Appendix~A for completeness. 

We consider a background universe with density parameter 
$\Om_0=\Om_m+\Om_\La=1$,
consisting of photons, cold dark matter (CDM),  baryons and
neutrinos. At very early times $z\gg z_{dec}\sim 1100$, photons and 
baryons form a perfectly coupled ideal fluid.
 As time evolves, and as the electron density drops due to recombination
of primordial helium and hydrogen, Compton
scattering becomes less frequent and higher moments in the photon
distribution develop. This epoch has to be described by a Boltzmann
equation. Long after recombination, free electrons are so sparse that
the collision term can be neglected, and photons evolve according
to the collisionless Boltzmann or Liouville equation. During the epoch of
interest here, neutrinos are always collisionless and thus obey the
Liouville equation. 

In the next subsection, we parameterize in a completely general
way the degrees of freedom of the seed energy momentum
tensor. Subsection~B is devoted to the perturbation of Einstein's
equations and the fluid equations of motion. Next we treat the Boltzmann
perturbation equation. In Subsection~D we explain how we
determine the power spectra of CMB anisotropies, density fluctuation
and peculiar velocities by means of the derived perturbation
equations and the unequal time correlators of the seed energy momentum
tensor, which are obtained by numerical simulations. In Subsection E
we give the initial conditions and a brief description of our
Boltzmann code.

\subsection{The seed energy momentum tensor}

Since the energy momentum tensor of the seeds, $\Th_{\mu\nu}$, has no
homogeneous background contribution, it is gauge invariant by itself
according to the Stewart-Walker Lemma\cite{StW}.

$\Th_{\mu\nu}$  can be calculated by solving the matter equations for
the seeds in the Friedmann {\em background} geometry (Since $\Th_{\mu\nu}$
has no background component it satisfies the unperturbed
``conservation'' equations.). We decompose
$\Th_{\mu\nu}$ into scalar, vector and tensor contributions.
They decouple within linear perturbation theory and it is thus  possible
to write  the equations for each of these contributions separately.
As always (unless noted otherwise), we work in Fourier space.
We parameterize the scalar $(S)$ vector $(V)$ and tensor
$(T)$ contributions to $\Th_{\mu\nu}$ in the form
\bea  \Th_{00}^{(S)} &=& M^2f_{\rho}
     \label{3seed00} \\
     \Th_{j0}^{(S)}  &=& iM^2k_jf_v 
     \label{3seed0j} \\
   \Th_{jl}^{(S)}   
    &=& M^2\left[(f_p + {1\over 3}k^2 f_{\pi})\de_{jl} - k_jk_lf_\pi\right]
  \label{3seedjl}\\
     \Th_{j0}^{(V)}  &=& M^2w^{(v)}_j       \\
   \Th_{jl}^{(V)}    &=&iM^2\frac{1}{2}\l(k_jw^{(\pi)}_l+k_lw^{(\pi)}_j\r) \\
   \Th_{jl}^{(T)}  &=& M^2\tau^{(\pi)}_{ij}  ~.
\eea
Here $M$ denotes a
typical mass scale of the seeds. In the case of topological defects we
set $M=\eta$, where $\eta$ is the symmetry breaking scale\cite{Review}.
The vectors $\bm w^{(v)}$ and $\bm w^{(\pi)}$ are transverse and
$\tau^{(\pi)}_{ij}$ is a transverse traceless tensor,
\[ \bk\cd\bm w^{(v)} = \bk\cd\bm w^{(\pi)}=
	k^i\tau^{(\pi)}_{ij}=\tau^{(\pi)\;j}_j = 0 ~.\]

From the full energy momentum tensor $\Th_{\mu\nu}$ which may
 contain scalar, vector and tensor contributions, the scalar parts
$f_v$ and $f_{\pi}$ of a given Fourier mode are determined by 
\[ ik^j\Th_{0j} = -k^2M^2f_v  ~, \]
\[ -k^ik^j(\Th_{ij} - \frac{1}{3}\de_{ij}\de^{kl}\Th_{kl}) =
   \frac{2}{3}k^4M^2f_{\pi}   \; . \]
On the other hand $f_v$ and $f_{\pi}$ are also
determined in terms of $f_{\rho}$ and $f_p$ by energy and momentum 
conservation,
\be \dot{f}_{\rho} +k^2f_v + {\dot{a}\over a}(f_{\rho} +3f_p) = 0 
\label{ec}~, \ee
\be \dot{f}_v + 2 {\dot{a}\over a}f_v - f_p +{2\over 3}k^2f_{\pi} = 0 ~.
 \label{3f}  \ee
Once $f_v$ is known it is  easy to extract $M^2w^{(v)}_j = 
 \Th_{0j} - ik_jM^2f_v$. For $w^{(\pi)}_i$ we use
\[ ik^j(\Th_{lj}- \Th_{lj}^{(S)}) = -k^2M^2w^{(\pi)}_l~.\]
Again,  $w^{(\pi)}_l$ can also be obtained in terms of  $w^{(v)}_l$ by
means of momentum conservation,
\be \dot{w}^{(v)}_l +2({\dot{a}\over a})w^{(v)}_l +
	{1\over 2}k^2w^{(\pi)}_l = 0 ~.\ee

The geometry perturbations  induced by the seeds are
characterized by the Bardeen potentials, $\Phi_s$ and $\Psi_s$, for
scalar perturbations, the potential for the shear of the extrinsic
curvature, $\bm\Si^{(s)}$, for vector perturbations and the gravitational wave
amplitude, $H_{ij}^{(s)}$, for tensor perturbations. Detailed
definitions of these variables and their geometrical interpretation
are given in Ref.~\cite{Review} (see also Appendix~A). Einstein's
equations link the seed perturbations of the geometry to the energy
momentum tensor of the seeds. Defining the dimensionless small
parameter 
\be
\ep\equiv 4\pi GM^2~, 
\ee
we obtain
\bea
k^2\Phi_s &=&\ep(f_\rho+3{\dot{a}\over a}f_v)  \label{Phis}\\
 \Phi_s +\Psi_s &=& -2\ep f_{\pi}  \label{Psis}  \\
 -k^2\Si^{(s)}_i &=& 4\ep w^{(v)}_i  \label{Sis}\\
\ddot{H}^{(s)}_{ij} +2 {\dot{a}\over a}\dot{H}^{(s)}_{ij} +
  k^2H^{(s)}_{ij} &=&
  2\ep\tau^{(\pi)}_{ij} ~. \label{Hs}
\eea
Eqs.~(\ref{Phis}) to (\ref{Hs}) would determine the geometric
perturbations  if the cosmic fluid were
perfectly unperturbed. In a realistic situation, however,
we have to add the fluid perturbations in
the geometry which are defined in the next subsection.
Only the total geometrical perturbations are
determined via Einstein's equations. In this sense, Eqs.~ (\ref{Phis}) 
to (\ref{Hs}) should be regarded as definitions for
$\Phi_s~,\Psi_s~,{\bf\Si}^{(s)}$ and $H^{(s)}_{ij}$.

A description of the numerical calculation of the energy momentum tensor of the
seeds for global texture is given in Section~III.

\subsection{Einstein's equations and the fluid equations}
\subsubsection{scalar perturbations}
Scalar perturbations of the geometry have two degrees of freedom which
can be cast in terms of the gauge-invariant Bardeen
potentials,  $\Psi$ and $\Phi$~ \cite{Ba,KS}. For Newtonian forms of matter
$\Psi=-\Phi$ is nothing else than the Newtonian gravitational
potential. For matter with significant anisotropic stresses, $\Psi$ and $-\Phi$
differ. In geometrical terms, the former represents the lapse function
of the zero-shear hyper-surfaces while the latter is a measure of
their 3-curvature\cite{Review}. In the presence of seeds, the
Bardeen potentials are given by
\bea 
\Psi &=&\Psi_s +\Psi_m~,  \label{dec1}\\
\Phi &=&\Phi_s +\Phi_m ~, \label{dec2}
\eea
where the indices $_{s,m}$ refer to contributions from
 a source (the seed)  and the cosmic fluid respectively.
The seed Bardeen potentials are given in Eqs.~(\ref{Phis}) and (\ref{Psis}).

To describe the scalar perturbations of the energy momentum tensor of a
given matter component, we use the variables  $D_g$, a
gauge-invariant variable for density fluctuations, $V$, the
potential of peculiar velocity fluctuations, and $\Pi$, a potential
for anisotropic stresses (which vanishes for CDM and baryons). A
definition of these variables in terms of the components of the 
energy momentum tensor of the fluids and the metric perturbations can
be  found in Refs.~\cite{KS} or \cite{Review} and in Appendix~A. 

Subscripts and superscripts $_\ga$, 
$_c$, $_b$ or $_\nu$ denote the radiation, CDM, baryon or neutrino 
fluids respectively.

Einstein's equations yield the following relation for the
matter part of the Bardeen potentials\cite{DS}

\bea
\Phi_m &=& {4\pi Ga^2\over k^2}\big[\rho_\ga D _g^{(\ga)}+
\rho_c D _g^{(c)} + \rho_b D _g^{(b)}+ \nonumber\\
&& \rho_\nu D _g^{(\nu)} -\{4\rho_\ga +3\rho_c +3\rho_b
+4\rho_\nu\}\Phi  \nonumber\\
&&  +3{\dot a\over a}k^{-1}
 \{{4\over 3}\rho_\ga V_\ga+\rho_c V_c   \nonumber\\ 
&& +\rho_bV_b
+{4\over 3}\rho_\nu V_\nu\}\big] \label{Phm}\\
\Psi_{m} &=&-\Phi_{m} - {8\pi Ga^2\over k^2}\left(p_\ga\Pi_\ga
+ p_\nu\Pi_\nu\right)~. \label{Psm}
\eea
Note the appearance of $\Phi=\Phi_s+ \Phi_m$ on the r.h.s. of Eq.~(\ref{Phm}).
Using the decompositions (\ref{dec1},\ref{dec2}) we can solve for $\Phi$
and $\Psi$ in terms of the fluid variables and the seeds. With the
help of Friedmann's equation, Eqs.~(\ref{Phm}) and~(\ref{Psm}) can
then be written in the form
\bea
\Phi &=& {1\over {2\over 3}\left({\dot{a}\over a}\right)^{-2}k^2
		{\small +4x_\ga +3x_c
  +3x_b+4x_\nu}}\big[  \nonumber \\ && x_\ga\si_0 + x_cD_g^{(c)}
+xD_g^{(b)} + x_\nu\nu_0 + \nonumber \\
&& +{\dot{a}\over
a}k^{-1}\left(4x_\ga V_\ga+3x_cV_c+3x_bV_b\right.  \nonumber\\ 
&& \left.  +4x_\nu V_\nu\right)
+{2\over 3}k^2\left({\dot{a}\over a}\right)^{-2}\Phi_s \big] \label{Phi}\\
\Psi &=&-\Phi-2\ep f_\pi- \nonumber \\
 && \left({\dot{a}\over a}\right)^2k^{-2}(x_\ga\Pi_\ga
+x_\nu\Pi_\nu)~. \label{Psi} 
\eea

Here we have normalized the
scale factor such that $a=1$ today. The density parameters
$\Om_{\small\bullet}$ always represent the values of the corresponding
density parameter today (Here $_{\small\bullet}$ stands for
$_c~,~_\ga~,~_b$ or $_\nu$.). To avoid
any confusion, we have introduced the variables $x_{\small\bullet}$
for the time dependent density parameters,
\bea
x_{\ga,\nu} &=& {\Om_{\ga,\nu}\over \Om_\ga + \Om_c a
+ \Om_b a+ \Om_\nu  +\Om_\La a^4} \\
x_{c,b} &=& {\Om_{c,b}a\over \Om_\ga + \Om_c a
+ \Om_b a + \Om_\nu +\Om_\La a^4}~.
\eea 

The fluid variables of photons and neutrinos are obtained by
integrating the scalar brightness perturbations, which we denote by
$\MM_S(t,\bk,\bn)$ and $\NN_S(t,\bk,\bn)$ respectively, over directions,
$\bn$,
\bea
D _g^{(\ga)} &=& {1\over 4\pi}\int \MM_S d\Om = \si_0 \label{ga0}\\
 V_\ga    &=& {3i\over 16\pi k}\int (\bk\cd\bn)\MM_S d\Om \\
 &=& {3\over 4}\si_1^{(S)}  \label{ga1}\\
\Pi_\ga   &=& {-9\over 8\pi k^2}\int ((\bk\cd\bn)^2-{1\over 3}k^2)\MM_S
 d\Om \\
 &=& 3 \si_2^{(S)} \label{ga2}\\
D _g^{(\nu)} &=& {1\over 4\pi}\int \NN_S d\Om =\nu_0 \label{n0}\\
 V_\nu    &=& {3i\over 16\pi k}\int (\bk\cd\bn)\NN_S d\Om \\
 &=& {3\over 4}\nu_1^{(S)} \label{n1} \\
\Pi_\nu   &=&{-9\over 8\pi k^2}\int ((\bk\cd\bn)^2-{1\over 3}k^2)\NN_S
 d\Om \\
  &=& 3 \nu_2^{(S)}  \label{n2}~.
\eea
A systematic definition of the modes $\si_j$ and $\nu_j$ is given
in the next subsection.

The equation of motion for CDM is given by energy and momentum
conservation,
\bea
\dot{D}_g^{(c)} +kV_c &=& 0~, \label{Dc}\\ 
\dot{V}_c +\left({\dot{a}\over a}\right)V_c &=& k\Psi ~. \label{Vc}
\eea
During the very tight coupling regime, $z\gg z_{dec}$, we may neglect
the baryon contribution in the energy momentum conservation of the
baryon-photon plasma. We then have
\bea
\dot{D}_g^{(\ga)} +{4\over 3}kV_\ga &=& 0~, \label{Dga}\\
\dot{V}_\ga -k{1\over 4}D_g^{(\ga)} &=& k(\Psi-\Phi)~,\label{Vga}\\
D_g^{(b)} &=& {3\over 4}D_g^{(\ga)}  \label{Dbt}\\
V_b &=& V_\ga ~. \label{Vbt}
\eea
The conservation equations for neutrinos are not very useful,
since they involve anisotropic stresses and thus do not close. At the
temperatures of interest to us, $T\ll1$MeV, 
neutrinos have to be evolved by means of the Liouville equation which
we discuss in the next subsection.

Once the baryon contribution to the baryon-photon fluid becomes
non-negligible, and the imperfect coupling of photons and baryons has
to be taken into account (for a 1\% accuracy of the results, the
redshift corresponding to this epoch is around $z\sim 10^7$), we
evolve also the photons with a Boltzmann equation. The equation of
motion for the baryons is then
\bea
\dot{D}_g^{(b)} +kV_b &=& 0~, \label{Db}\\ 
\dot{V}_b +\left({\dot{a}\over a}\right)V_b &=& k\Psi
-{4\si_Tn_e\Om_\ga \over 3\Om_b}[V_\ga-V_b]~. \label{Vb}
\eea
The last term in Eq.~(\ref{Vb}) represents the photon drag force
induced by non-relativistic Compton scattering, $\si_T$ is the
Thomson cross section, and $n_e$ denotes the number density of free electrons.
At very early times, when $\si_Tn_e\gg 1/t$, the 'Thomson drag' just
forces $V_b=V_\ga$, which together with Eqs.~(\ref{Dga}) and
(\ref{Db}) implies (\ref{Dbt}).

An interesting phenomenon often called 'compensation' can be important
on super horizon scales, $kt\ll 1$. If we neglect anisotropic stresses
of photons and neutrinos and take into account that ${\cal
O}(D_g) = {\cal O}(ktV)$ and ${\cal O}(V) = {\cal O}(kt\Psi)$ for
$kt\ll 1$, Eqs.~(\ref{Phi}) and (\ref{Psi}) lead to
\be
 {\cal O}(\Phi) =  {\cal O}\left((kt)^2\Phi_s -2\ep f_\pi\right) 
	~. \label{comp}
\ee
Hence, if anisotropic stresses are relatively small, $\ep f_\pi\ll \Phi_s$,
the resulting gravitational potential on super horizon scales is much
smaller than the
one induced by the seeds alone. One must be very careful not to over
interpret this 'compensation' which is by no means related to
causality, but is due to the initial condition $D_g~,V\ra_{t\ra 0}0$. 
A thorough discussion of this issue is found in
Refs.~\cite{DS,MC,UDT}. As we shall see in the next section, for textures 
$\Phi_s$ and $\ep f_\pi$ are actually of the same order. Therefore
Eq.~(\ref{comp}) does not lead to compensation, but it indicates
that CMB anisotropies on very large scales (Sachs-Wolfe effect) are
dominated by the amplitude of seed anisotropic stresses.
\vspace{13pt}

The quantities which we want to calculate and compare with
observations are the CDM density power spectrum and the peculiar
velocity power spectrum today
\bea
 P(k) &=& \langle|D_g^{(c)}(k,t_0)|^2\rangle \\
\mbox{and} && \nonumber\\
P_v(k) &=& \langle|V_c(k,t_0)|^2\rangle ~.
\eea
Here $\langle\cdots\rangle$ denotes an ensemble average over models.
Note that even though $D_g$ and $V$ are gauge invariant quantities
which do not agree with, {\em e.g.}, the corresponding quantities in
synchronous gauge, this difference is very small on subhorizon scales
(of order $1/kt$) and can thus be ignored.

On subhorizon scales the seeds decay, and CDM perturbations evolve
freely. We then have, like in inflationary models,
\be P_v(k) = H_0^2\Om_m^{1.2}P(k)k^{-2} ~.\ee

\subsubsection{vector perturbations}
 Vector perturbations of the geometry have two degrees of freedom
which can be cast in a divergence free vector field. A gauge-invariant
quantity describing vector perturbations of the geometry is $\bm\Si$, a
vector potential for the shear tensor of  the $\{t=$const.$\}$
hyper-surfaces. Like for scalar perturbations, we split the
contribution to $\bm\Si$ into a source term  coming from the seeds
given in the previous subsection, and a
part due to the vector perturbations in the fluid,
\be \bm\Si = \bm\Si_s + \bm\Si_m  \label{Si} ~.\ee
The perturbation of Einstein's equation for
${\bm\Si}_m$ is\cite{Review}
\bea
k^2\bm\Si_m &=& 6\left({\dot{a}\over a}\right)^2[{4\over
3}x_\ga\bm\om_\ga + x_c\bm\om_c + \nonumber \\ 
 && + x_b\bm\om_b + {4\over 3}x_\nu\bm\om_\nu] \label{Sim}~.
\eea
Here $\bm\om_{\small\bullet}$ is the fluid vorticity which
generates the vector type shear of the equal time hyper-surfaces
(see Appendix~A). By definition, vector perturbations are transverse,
\be \bm\Si\cd\bk= \bm\Si_m\cd\bk= \bm\Si_s\cd\bk=
\bm\om_{\small\bullet}\cd\bk = 0~. \label{vector}
\ee

It is interesting
to note that vector perturbations in the geometry do not induce any
vector perturbations in the CDM (up to unphysical gauge modes), since
no geometric terms enter the momentum conservation for CDM vorticity,
\[ \dot{\bm\om}_c + {\dot{a}\over a}\bm\om_c = 0 ~,\]
hence we may simply set $\bm\om_c = 0$.
This is also the case for the tightly coupled baryon radiation
plasma. But as soon as higher moments in the photon distribution build
up, they feel the vector perturbations in the geometry (see next
section) and transfer it onto the baryons
via the photon drag force, 
\be
\dot{\bm\om}_b +\left({\dot{a}\over a}\right)\bm\om_b =
{4\si_Tn_e\Om_\ga \over 3\Om_b}[\bm\om_\ga-\bm\om_b]~. \label{wb}
\ee
The photon vorticity is given by an integral over the vector
type photon brightness perturbation, $\MM_V$,
\be
\bm\om_\ga = \frac{1}{4\pi}\int \bn\MM_Vd\Om ~,
          \label{omga}  
\ee
where the integral is over  photon directions, $\bn$.
In terms of the development presented in the next section for $\bk$
pointing in $z$-direction, we obtain
\be
\bm\om_\ga =\left(\si_{1,2}^{(V)} +\si_{1,0}^{(V)}~,~ 
	\si_{2,2}^{(V)} +\si_{2,0}^{(V)}~,~ 0\right) ~.
\ee
Equivalently, we have for neutrinos
\bea
\bm\om_\nu &=& \frac{1}{4\pi}\int \bn\NN_Vd\Om ~,
          \label{omnu}   \\
\bm\om_\nu &=& \left(\nu_{1,2}^{(V)} +\nu_{1,0}^{(V)}~,~ 
	\nu_{2,2}^{(V)} +\nu_{2,0}^{(V)}~,~ 0\right) ~.
\eea
The vector equations of motion for photons and neutrinos are discussed
in the next section.

\subsubsection{tensor perturbations}
Metric perturbations also have  two tensorial degrees of freedom,
gravity waves, which
are represented by the two helicity states of a transverse traceless
tensor (see Appendix~A). As before, we split the geometry perturbation
into a part induced by the seeds and a part due to the matter fluids,
\be
 H_{ij} = H^{(s)}_{ij} +H^{(m)}_{ij} \label{Hij} ~.
\ee
The only matter perturbations which generate gravity waves are tensor
type anisotropic stresses which are present in the photon and neutrino
fluids. The perturbation of Einstein's equation yields
\bea
&&\ddot{H}^{(m)}_{ij} +2\left({\dot{a}\over a}\right)\dot{H}^{(m)}_{ij} 
 +k^2H^{(m)}_{ij} = \nonumber \\
&&\left({\dot{a}\over a}\right)^2(x_\ga\Pi_{ij}^{(\ga)} +
x_\nu\Pi_{ij}^{(\nu)}) ~.
\eea

The relation between the tensor brightness perturbations $\MM_T$,
$\NN_T$ and the tensor anisotropic stresses, $\Pi_{ij}^{(\ga)}$ and
$\Pi_{ij}^{(\nu)}$ is given by
\bea
\Pi^{(\ga)}_{ij} &=&{3\over 4\pi}\int(n_in_j - {1\over 3}\de_{ij})
	\MM_T d\Om~,    \label{Piga} \\
\Pi^{(\nu)}_{ij} &=&{3\over 4\pi}\int(n_in_j - {1\over 3}\de_{ij})
	\NN_T d\Om~.    \label{Pinu} ~
\eea
In terms of the development presented in the next section for $\bk$
pointing in $z$-direction, we have
\bea
\Pi^{(\ga)}_{11} &=& -\Pi^{(\ga)}_{22} = {6\over 35}\si_{+,4} +
	  {4\over 7}\si_{+,2}+ {2\over 5}\si_{+,0}~,  \\
\Pi^{(\ga)}_{12} &=& \Pi^{(\ga)}_{21} = {6\over 35}\si_{\times,4} +
	  {4\over 7}\si_{\times,2}+ {2\over 5}\si_{\times,0}~,  \\
\Pi^{(\nu)}_{11} &=& -\Pi^{(\nu)}_{22} = {6\over 35}\nu_{+,4} +
	  {4\over 7}\nu_{+,2}+ {2\over 5}\nu_{+,0}~,  \\
\Pi^{(\nu)}_{12} &=& \Pi^{(\nu)}_{21} = {6\over 35}\nu_{\times,4} +
	  {4\over 7}\nu_{\times,2}+ {2\over 5}\nu_{\times,0}~.
\eea
We find that the effect of anisotropic stresses of photons and
neutrinos is less than 1\% in the final result, and hence we have
neglected them.
 
\subsection{The Boltzmann equation}
When particle interactions are less frequent, the fluid
approximation is not  sufficient, and we have to describe the
given particle species by a Boltzmann equation, in order to take into account
phenomena like collisional and directional dispersion. In the case of
massless particles like massless neutrini or photons, the Boltzmann
equation can be integrated over energy, and we obtain an
equation for the brightness perturbation which depends only on
momentum directions\cite{Review}.  As before, we
split the brightness perturbation into a scalar, vector and
tensor component, and we discuss the perturbation equation of each
of them separately,\footnote{We could in principle add higher 
spin components to the distribution functions. But they are not 
seeded by gravity and
since photons and neutrinos interact at high enough temperatures, they
are also absent in the initial conditions.} 
\bea
\MM  &=&\MM_S +\MM_V +\MM_T \\
\mbox{and} &&\\
\NN  &=&\NN_S +\NN_V +\NN_T.
\eea
The functions $\MM$ and $\NN$ depend on the wave vector $\bk$, the
photon (neutrino) direction $\bn$ and conformal time $t$.
Linear polarization of photons induced by Compton scattering
is described by the variable $\MM^{(Q)}$ (the Stokes parameter $Q$)
depending on the same variables.
We choose for each $\bk$-mode a reference system with 
$z$-axis parallel to $\bk$. 
For scalar perturbations we achieve in this way 
azimuthal symmetry --- the left hand side of the Boltzmann equation and
therefore also the brightness $\MM$ depend only
on $\mu =  (\hat{\bk}\cd \bn)$ and can be developed
in Legendre polynomials. 

The left hand side of the Boltzmann equation for vector and tensor
perturbations also determines the azimuthal dependence of $\MM$ for
vector and tensor perturbations, as we shall see in detail.

\subsubsection{scalar perturbations}

We expand the brightness $\MM_S(\bk,\bn,t)$
in the form:

\be
\MM_S(\bn,\bk,t) = \sum_{\ell=0}^\infty
 (-i)^{\ell} (2 \ell +1) \si^{(S)}_{\ell}(t,\bk)P_{\ell}(\mu),\label{expMM}
\ee

where $P_\ell$ denotes the Legendre
polynomial of order $\ell$ and $\si^{(S)}_{\ell}$ is the associated
multipole moment.
An analogous decomposition also applies to the amplitude of polarization
anisotropy, $\MM_S^Q(\bn,\bk,t)$, and we denote
 the associated multipole moment by  $q^{(S)}_{\ell}$.  

The Boltzmann equation for scalar perturbations in the photon
brightness and polarization is~\cite{Review,melvit}

\bea
\lefteqn{\dot\MM_S +i\mu k \MM_S = 4i\mu k (\Phi-\Psi) +} \nonumber \\
 && a\si_Tn_e[ D_g^{(\ga)}
-\MM_S- 4i\mu V_b -{1 \over 2}P_2(\mu)Q]~,
\label{Mdot}
\eea
\bea
\lefteqn{\dot\MM_S^{(Q)} +i\mu k \MM_S^{(Q)} =} \nonumber \\
 && a\si_Tn_e[
-\MM_S^{(Q)} +{1 \over 2}\big(1-P_2(\mu\big))Q]~,
\label{MdotQ}
\eea
where
\bean Q=\sigma_2^{(S)}+q^{(S)}_0+q^{(S)}_2~.
\eean

The first term on the right hand side of Eq.~(\ref{Mdot}) represents 
the gravitational
interaction (photons without collisions move along lightlike geodesics of
the perturbed geometry), while the term in square brackets is the collision
integral for non-relativistic Compton scattering.

Inserting expansion (\ref{expMM}) into Eqs.~(\ref{Mdot}) and (\ref{MdotQ})
using the standard recursion relations for Legendre polynomials, we
obtain the following series of coupled equations:

\be
\dot{\si}^{(S)}_0 +{k}\si^{(S)}_1 = 0~, \label{si0}
\ee
\bea
\lefteqn{\dot{\si}^{(S)}_1 -{k \over 3}[\si^{(S)}_0-2\si^{(S)}_2] =} \nonumber \\
 && {4 \over 3} k (\Psi-\Phi)
 +a\si_Tn_e[{4 \over 3}V_b-\si^{(S)}_1] ~, \label{si1}
\eea
\bea
\lefteqn{\dot\si^{(S)}_2 -{k \over 5}[ 2 \si^{(S)}_1 -
3 \si^{(S)}_3] =} \nonumber \\
&& -a\si_Tn_e[\si^{(S)}_2 -{1\over 10}Q] ~,   \label{si2} 
\eea
\bea
\dot{\si}^{(S)}_\ell -  &{k \over {2 \ell +1}}
	\left[\ell\si^{(S)}_{\ell-1} -
   (\ell+1)\si^{(S)}_{\ell+1}\right] &= \nonumber \\
 -a\si_Tn_e\si^{(S)}_\ell~, &\!\! \mbox{for }~~ \ell\ge 3~.&\label{sil}
\eea
and
\bea
\dot{q}^{(S)}_\ell -{k \over {2 \ell +1}}
	\left[\ell q^{(S)}_{\ell-1} -
   {(\ell+1)}q^{(S)}_{\ell+1}\right] &=& \nonumber \\
 +a\si_Tn_e[-q^{(S)}_\ell+
{1 \over 2}Q(\delta_{\ell 0} +{1\over 5}\delta_{\ell 2})]. && \label{silQ}
\eea

For the neutrinos we obtain the same equations just without
collision integral

\be
\NN_S(\bn,\bk,t) = \sum_{\ell=0}^\infty 
 (-i)^{\ell} (2 \ell +1) \nu^{(S)}_{\ell}(t,\bk)P_{\ell}(\mu),\label{expNN}
\ee
and
\bea
 &\dot{\nu}^{(S)}_\ell -{k \over {2 \ell +1}}
	\left[\ell\nu^{(S)}_{\ell-1} -
   {(\ell+1)}\nu^{(S)}_{\ell+1}\right] &=  \nonumber \\
&& {4 \over 3} k(\Psi-\Phi)\delta_{\ell 1}.\label{siln}
\eea

We are interested in the power spectrum of CMB anisotropies
which is defined by
\be \left\langle{\delta T\over T}({\bf n}){\delta T\over T}({\bf n}')
\right\rangle\left|_{{~}_{\!\!({\bf n\cdot n}'=\cos\vartheta)}}\right. =
  {1\over 4\pi}\sum_\ell(2\ell+1)C_\ell P_\ell(\cos\vartheta)~. \ee

Here $\langle\cdots\rangle$ denotes the ensemble average over models.
We assume that an 'ergodic hypothesis' is satisfied and we can
interchange spatial and ensemble averages. The problem that actual
observations can average at best over one horizon volume is known under
the name 'cosmic variance'. It  severely restricts the accuracy with
which, for example, low multipoles of CMB anisotropies observed in our
horizon volume can be predicted for a given model.
 
Using the addition theorem of spherical harmonics, one obtains, with the
Fourier transform conventions adopted here, (for details see
Appendix~B)
\be
C_\ell^{(S)} =  {1 \over {8 \pi}}\int
 k^2dk\langle|\si^{(S)}_\ell(t_0,k)|^2\rangle ~,
\label{ClS}\ee
where the superscript $^{(S)}$ indicates that Eq.~(\ref{ClS}) gives
the contribution from {\em scalar} perturbations.

\subsubsection{vector perturbations}
 Vector perturbations  are very small on angular scales
corresponding to $\ell\gsim 500$, where Compton scattering and thus
polarization become relevant. We therefore neglect polarization in this case.
The Boltzmann equation for vector perturbations then reads
\bea
\lefteqn{\dot\MM_V +i\bk\cd\bn\MM_V = -4i(\bn\cd\bk)(\bn\cd\bm\Si) 
	+} \nonumber \\
 && a\si_Tn_e[ 4(\bn\cd\bm\om_b) -\MM_V +{1\over 2}n_{ij}M_{ij}]~,
\label{MVdot}
\eea
where
\bean n_{ij} &\equiv& n_in_j-{1\over 3}\de_{ij} ~~\mbox{ and} \\
 M_{ij} &=& {3\over 8\pi}\int n_{ij}\MM_V d\Om ~.
\eean

We use coordinates for which $\bk$ is parallel to the $z$-axis. Then
\[ \bm\Si= (\Si_1,\Si_2,0)~~,~~~ \bm\om=
(\om_1,\om_2, 0) \]
 and 
\[ \bn=(\sqrt{1-\mu^2}\cos\vph,
	\sqrt{1-\mu^2}\sin\vph,\mu)~.\]

With the ansatz 
\bea \MM_V(\bk,\bn,t) &=& \sqrt{1-\mu^2}[\MM_1^{(V)}(\bk,\mu,t)\cos\vph 
	+ \nonumber \\
&& +\MM_2^{(V)}(\bk,\mu,t)\sin\vph]~, \label{M12}
\eea
the equations for $\MM_{1,2}$  decouple and 
 the right hand side of Eq.~(\ref{MVdot}) depends only on $\mu$. 
Like for scalar 
perturbations, we expand $\MM_{1,2}$ in Legendre polynomials
\be
\MM_{\epsilon}^{(V)}(\mu,\bk,t) = \sum_{\ell=0}^\infty
 (-i)^{\ell} (2 \ell +1) \si^{(V)}_{\epsilon,\ell}(t,\bk)P_{\ell}(\mu),
\label{expMV}
\ee
where ${\epsilon} =1,2$.

Eq.~(\ref{MVdot}) then leads to
\bea
\lefteqn{\dot\MM^{(V)}_{\epsilon} +i\mu k\MM^{(V)}_{\epsilon} = 
-4i\mu k\Si_{\epsilon} +} \nonumber \\
&&a\si_Tn_e[4\om_\ep^{(b)} -\MM^{(V)}_\epsilon -i\mu 
{3 \over {10}}(\si_{{\epsilon},1}^{(V)}+\si_{{\epsilon},3}^{(V)})]~.
 \label{dotM1}
\eea
With Eq.~(\ref{expMV}), this can be expressed as the following set of
coupled equations for the variables $\si^{(V)}_{\ep,\ell}$.

\be
	\dot{\si}^{(V)}_{\ep,0} +k\si^{(V)}_{\ep,1} = 
	a\si_Tn_e[4\om_\ep^{(b)}-\si^{(V)}_{\ep,0}]~, \label{Vsi0}
\ee
\bea
\lefteqn{\dot{\si}^{(V)}_{\epsilon,1} -{k\over3}[\si^{(V)}_{\epsilon,0}
	-2\si^{(V)}_{\epsilon,2}] =} \nonumber \\
 	&&  +{4 \over 3} k\Si_{\epsilon}
-a\si_Tn_e[{9\over 10}\si^{(V)}_{\epsilon,1}-{1\over 10}\si^{(V)}_{\epsilon,3}]
 ~, \label{Vsi1}\\
&& \mbox{ and }\nonumber \\
\lefteqn{\dot{\si}^{(V)}_{\epsilon,\ell} -{k \over {2 \ell +1}}
	\left[\ell\si^{(V)}_{\epsilon,\ell-1} -
   {(\ell+1)}\si^{(V)}_{\epsilon,\ell+1}\right] =} \nonumber \\
&& -a\si_Tn_e\si^{(V)}_{\epsilon,\ell}~~\mbox{ for } \ell\ge 2.\label{Vsil}
\eea

For neutrino perturbations  we obtain the same equations  up
to the collision term. We repeat them here for completeness.
\bea
\dot{\nu}^{(V)}_{\epsilon,\ell} -{k \over {2 \ell +1}}
	\left[\ell\nu^{(V)}_{\epsilon,\ell-1} -
   {(\ell+1)}\nu^{(V)}_{\epsilon,\ell+1}\right]& =& \nonumber \\
& +{{4 \over 3}} \Si_{\epsilon} k \delta_{\ell 1}. &\label{Vsiln}
\eea

As for scalar perturbations, the CMB anisotropy power spectrum is 
obtained by integration over $k$-space. One finds (see Appendix~B),
\be
C_\ell^{(V)} = {\ell(\ell+1) \over {8 \pi}}  \int k^2dk{{\langle|
\si^{(V)}_{1,\ell+1}(t_0,k)+\si^{(V)}_{1,\ell-1}
(t_0,k)|^2\rangle}\over {(2 \ell+1)^2}} ~.
\label{ClV}\ee
Here the fact that there are two equal contributions from both
polarization states, $\ep=1,2$ (statistical isotropy) is
taken care of.

\subsubsection{tensor perturbations}
For tensor perturbations, and a wave vector $\bk$ pointing into
the $3$-direction,  the only non vanishing components
of the perturbed metric tensor are $H_{11}= - H_{22}=H_{+}$ and
$H_{12}=H_{21}=H_{\times}$.
Neglecting polarization, the Boltzmann equation for tensor 
perturbations is\cite{Review}
\bea
 \lefteqn{\dot{\MM}_T +ik\mu\MM_T = -4n^in^j\dot{H}_{ij}} \nonumber \\
 &&-a\si_Tn_e[\MM_T-{1\over 2}n_{ij}M_{ij}] \label{MTdot}~.
\eea
With the ansatz
\bea
 \MM_T(\bk,\bn,t) &=& (1-\mu^2)[\MM^{(T)}_{+}(\bk,\mu,t)\cos 2 \vph 
	+ \nonumber \\
&& +\MM^{(T)}_{\times}(\bk,\mu,t)\sin 2 \vph]~, \label{M+x}
\eea
the two modes $\MM^{(T)}_{+,\times}$ decouple completely and the right
hand side of Eq.~(\ref{MTdot}) depends only on $\mu$. We can then 
expand the modes in terms of  Legendre polynomials
\be
\MM^{(T)}_{\epsilon}(\mu,k,t) = \sum_{\ell=0}^\infty
 (-i)^{\ell} (2 \ell +1) \si^{(T)}_{{\epsilon},\ell}(t,\bk)P_{\ell}(\mu),
\label{expMT}
\ee
where ${\epsilon}= +,\times$. Eq.~(\ref{MTdot}) now becomes
\bea
\lefteqn{\dot\MM^{(T)}_{\epsilon} +i\mu k\MM^{(T)}_{\epsilon} = 4 {\dot H}_{\epsilon}
 +} \nonumber \\
&& +a\si_Tn_e[-\MM^{(T)}_{\epsilon} + {1 \over 10}\si^{(T)}_{{\epsilon},0}
 + {1 \over 7}\si^{(T)}_{{\epsilon},2}
+{3 \over 70}\si^{(T)}_{{\epsilon},4}]~,
\eea
leading to the series of coupled equations for the coefficients 
$\si^{(T)}_{\ep,\ell}$
\bea
\lefteqn{ \dot{\si}^{(T)}_{\epsilon,0} +k\si^{(T)}_{\epsilon,1} = 
4{\dot H}_{\epsilon} +}  \nonumber \\
 && + a\si_Tn_e[ -{9\over 10}\si^{(T)}_{\epsilon,0} 
 + {1 \over 7 }\si^{(T)}_{\epsilon,2}
 +{3 \over 70}\si^{(T)}_{\epsilon,4}]~,  \label{Tsi0}\\
\lefteqn{\dot{\si}^{(T)}_{\epsilon,\ell} -{k \over 2\ell +1}
	\left[\ell\si^{(T)}_{\epsilon,\ell-1} -
   (\ell+1)\si^{(T)}_{\epsilon,\ell+1}\right] =} \nonumber \\   &&
  -a\si_Tn_e\si^{(T)}_{\epsilon,\ell},~ \mbox{ for } \ell\ge 1~.\label{Tsil}
\eea

As before, the CMB anisotropy power spectrum is obtained by integration
over $k$-space (see Appendix~B),
\be
C_\ell^{(T)} = {1 \over {8 \pi}} {{(\ell+2)!} \over {(\ell-2)!}}
 \int k^2dk{{\langle|
\Si^{(T)}_{\ell}|^2\rangle}\over {(2 \ell+1)^2}} ~,
\label{ClT}\ee
where
\be
\Si^{(T)}_{\ell}={{\si^{(T)}_{{\epsilon},\ell-2}} \over {2 \ell -1}}
-{{2{(2 \ell +1)\si^{(T)}_{{\epsilon},\ell}}} \over {(2 \ell -1)(2 \ell +3)}}
+{{\si^{(T)}_{{\epsilon},\ell+2}} \over {{2 \ell +3}}}~.
\ee

\subsection{Eigenvector expansion of the source correlators}
In the previous subsections we have derived a closed system of linear
differential equations with source terms. The source terms are linear
combinations of the seed energy momentum tensor which is determined by
numerical simulations. A given realization of our model has
 random initial conditions; the  seed energy 
momentum tensor is a random variable. In principle we could
calculate the induced random variables $D_g^{(c)}(\bk,t_0)$,
$V_c(\bk,t_0)$, $\si^{(\small\bullet)}_\ell(\bk,t_0)$ etc for 100 to 
1000 realizations of our
model and determine the expectation values  $P(k)$, $P_v(k)$ and $C_\ell$ 
by averaging. This procedure has been adapted in Ref.~\cite{ABR} for a
seed energy momentum tensor modeled by a few random parameters.

In the case of a seed energy momentum tensor coming entirely from
numerical simulations, this procedure is  not feasible. The first and most
important bottleneck is the dynamical range of the simulations which
is about $40$ in our largest $(400)^3$ simulation,
taking around 5 hours CPU time on a NEC SX-4 supercomputer.
To determine the $C_\ell$'s for $2\le\ell\le 1000$ we need a dynamical
range of about 10'000 in $k$-space (this means $k_{\max}/k_{\min} \sim
10'000$, where $k_{\max}$ and $k_{\min}$ are the maximum and minimum wave
numbers which contribute to the $C_\ell$'s within our accuracy ( $\sim$
10\%).

With brute force, this problem is thus not tractable with present or
near future computing capabilities. But there are a series of theoretical
observations which reduce the problem to a feasible one:

For each wave vector $\bk$ given, we have to solve a system of 
linear perturbation equations with random sources,
\be
\DD X = \SS~. \label{lineq}
\ee
Here $\DD$ is a time dependent linear differential operator, 
$X$ is the vector of
our matter perturbation variables specified in the previous
subsections (photons, CDM, baryons and neutrini;
total length up to $2000$), and $\SS$ is the random source term, 
consisting of linear combinations of the seed energy momentum tensor.

For given initial conditions, this equation can be solved by
means of a Green's function (kernel), $\GG(t,t')$, in the form
\be
X_j(t_0,\bk) =\int_{t_{in}}^{t_0}\! dt\GG_{jl}(t_0,t,\bk)\SS_l(t,\bk)~.
\label{Gsol}
\ee
We want to compute power spectra or, more generally, quadratic 
expectation
values of the form
\[  \langle X_j(t_0,\bk)X_l^*(t_0,\bk)\rangle ~,\]
which, according to Eq.~(\ref{Gsol}) are given by
\bea
\lefteqn{\langle X_j(t_0,\bk)X_l^*(t_0,\bk)\rangle =} \nonumber \\
&& \int_{t_{in}}^{t_0}\! dt\GG_{jm}(t_0,t,\bk)
  \int_{t_{in}}^{t_0} \! dt'\GG^*_{ln}(t_0,t',\bk)\times \nonumber \\
&& \langle\SS_m(t,\bk)\SS_n^*(t',\bk)\rangle~. \label{power}
\eea
The only information about the source random variable which we really
need in order to compute power spectra are therefore the unequal time
 two point correlators
\be
\langle\SS_m(t,\bk)\SS_n^*(t',\bk)\rangle~. \label{2point}
\ee
This nearly trivial fact has been exploited by many workers in the 
field, for the first time probably in Ref.~\cite{ACFM} where the
decoherence of models with seeds has been discovered, and later in
Refs.~\cite{PST,Aetal,KD,DS} and others.

To solve the enormous problem of dynamical range, we make use
of 'scaling', statistical isotropy and causality. 

We call seeds 'scaling' if their correlation functions 
$C_{\mu\nu\rho\la}$ defined by
\bea 
\Th_{\mu\nu}(\bk,t) &=& M^2\th_{\mu\nu}(\bk,t) ~, \\
C_{\mu\nu\rho\la}(\bk,t,t') &=&\langle\th_{\mu\nu}(\bk,t)
	\th_{\rho\la}^*(\bk,t')\rangle \label{cor}
\eea
are scale free; {\em i.e.} the only  dimensional parameters in
$C_{\mu\nu\rho\la}$ are the variables  $t,t'$ and $\bk$ themselves. Up to a
certain number of dimensionless functions $F_n$ of $z=k\sqrt{tt'}$ and
$r=t/t'$, the correlation functions are then determined by the requirement
of statistical isotropy, symmetries and by their dimension. Causality requires the 
functions $F_n$ to be analytic in $z^2$. A more detailed investigation
of these arguments and their consequences is presented in
Ref.~\cite{DK}. There we show that statistical isotropy and energy
momentum conservation reduce the correlators ~(\ref{cor}) to five
such functions $F_1$ to $F_5$. 

In cosmic string simulations, energy and momentum are not
conserved. Strings loose their energy by radiation of gravitational
waves and/or massive particles. In this case 14 functions of $z^2$ and
$r$ are needed to describe the unequal time correlators\cite{CHM}.

Since analytic functions generically
are  constant for small arguments $z^2\ll 1$, $F_n(0,r)$ actually
determines $F_n$ for all values of $k$ with $z=k\sqrt{tt'}\lsim
0.5$. Furthermore, the correlation functions decay inside the horizon
and we can safely set them to zero for $z\gsim 40$ where they have
decayed by about two orders of magnitude (see Figs.~1 to 11). 
Making use of these generic
properties of the correlators, we have reduced the dynamical range
needed for our computation to about 40, which can be attained with
the $(256)^3$ to $(512)^3$ simulations feasible on present supercomputers. 

For the {\em scalar} part we need the correlators
\bea
  \langle\Phi_s(\bk,t)\Phi_s^*(\bk,t')\rangle &=&
	{1\over k^4\sqrt{tt'}}C_{11}(z,r)~, \label{Cs11}\\
  \langle\Phi_s(\bk,t)\Psi_s^*(\bk,t')\rangle &=& 
 	{1\over k^4\sqrt{tt'}}C_{12}(z,r)~, \label{Cs12}\\
  \langle\Psi_s(\bk,t)\Psi_s^*(\bk,t')\rangle &=& 
	{1\over k^4\sqrt{tt'}}C_{22}(z,r)~, \label{Cs22}
\eea
as well as $C_{21}(z,r) =C^*_{12}(z,1/r)$. The functions $C_{ij}$ 
are analytic in $z^2$. The pre-factor $1/(k^4\sqrt{tt'})$ comes from the
fact that the correlation functions $\langle f_\rho f^*_\rho\rangle$,  
$ k^4\langle f_\pi f^*_\pi\rangle$ and  $\langle f_v f^*_v\rangle$ have to
be analytic and from dimensional considerations (see Ref.~\cite{DK}). 

The functions $C_{ij}$ are shown
in Figs.~1 to~3. Panels~(a) are obtained from numerical
simulations.  Panels~(b) represent the same correlators for the
large-$N$ limit of global $O(N)$-models (see\cite{TSlN,KD}).
\begin{figure}[ht]
\centerline{\epsfig{figure=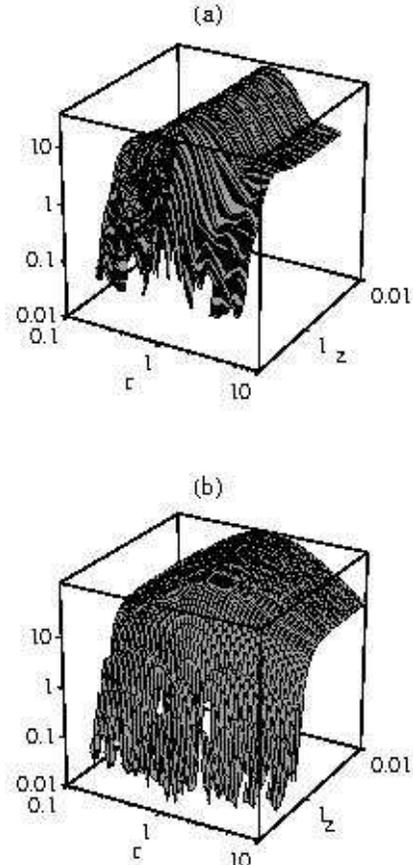,width=5.5cm}}
\caption{\label{fig1}The two point correlation function 
 $C_{11}(z,r) = 
 k^4\sqrt{tt'}\langle\Phi_s(\bk,t)\Phi_s^*(\bk,t')\rangle$ is
 shown. Panel (a) represents the result from numerical simulations of
 the texture model; panel (b) shows the large-$N$ limit.  For
 fixed $r$ the correlator is constant for $z<1$ and then decays. Note
 also the symmetry under $r\ra 1/r$.}
\end{figure}
\begin{figure}[ht]
\centerline{\psfig{figure=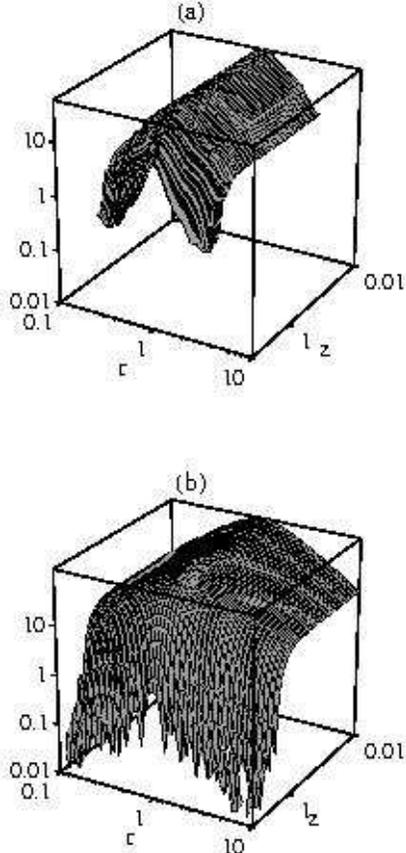,width=5.5cm}}
\caption{\label{fig2}The same as Fig.~\ref{fig1} but for 
$C_{22}(z,r) = 
 k^4\sqrt{tt'}\langle\Psi_s(\bk,t)\Psi_s^*(\bk,t')\rangle$.}
\end{figure}
\begin{figure}[ht]
\centerline{\psfig{figure=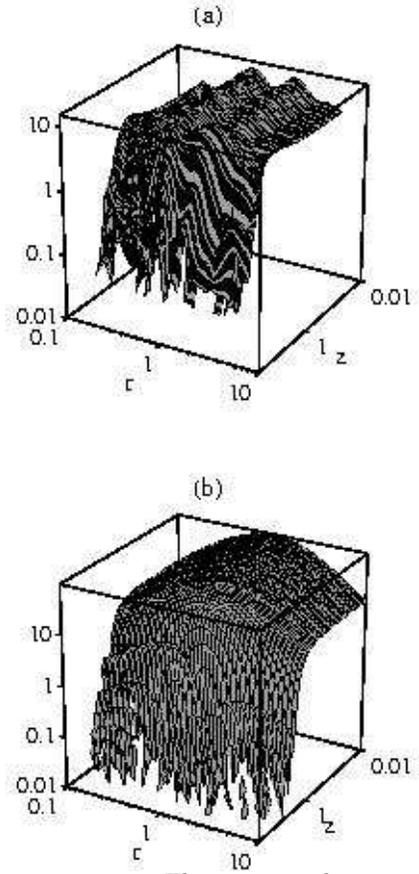,width=5.5cm}}
\caption{\label{fig3} The unequal time correlator, 
  $|C_{12}(z,r)| = 
 k^4\sqrt{tt'}|\langle\Phi_s(\bk,t)\Psi_s^*(\bk,t')\rangle|$ is
 shown. Note that the  $r\ra 1/r$ symmetry is lost in this case.}
\end{figure}

In Fig.~4 we show $C_{ij}(z,r=1)$, and in Fig.~5 the
'constant' of the Taylor expansion for $C_{ij}$ is given as a function
of $r$, {\em i.e.},  $C_{ij}(0,r)$.
\begin{figure}[ht]
\centerline{\psfig{figure=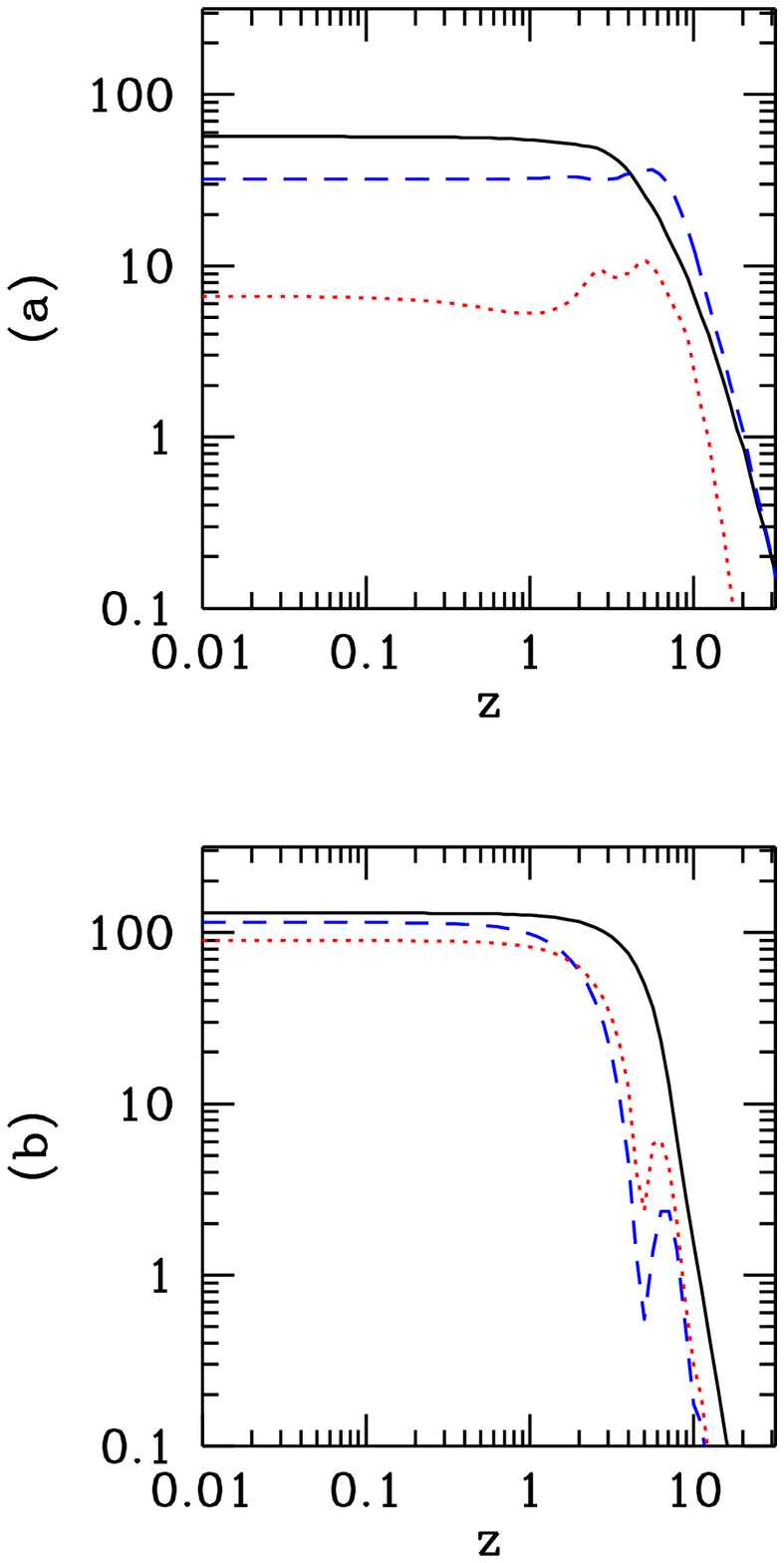,width=5cm}}
\caption{\label{fig4}
 The correlators  $C_{ij}(z,1)$ are
 shown. The solid, dashed and dotted lines represent $C_{22}~,~C_{11}$
 and $|C_{12}|$ respectively. Panel (a) is obtained from numerical
 simulations of the texture model and panel (b) shows the large-$N$
 limit. A striking difference is that the large-$N$ value for
 $|C_{12}|$ is relatively well approximated by the perfectly coherent
 result $\sqrt{|C_{11}C_{22}|}$ while the texture curve for $|C_{12}|$
 lies nearly a factor 10 lower.}
\end{figure}
\begin{figure}[ht]
\centerline{\psfig{figure=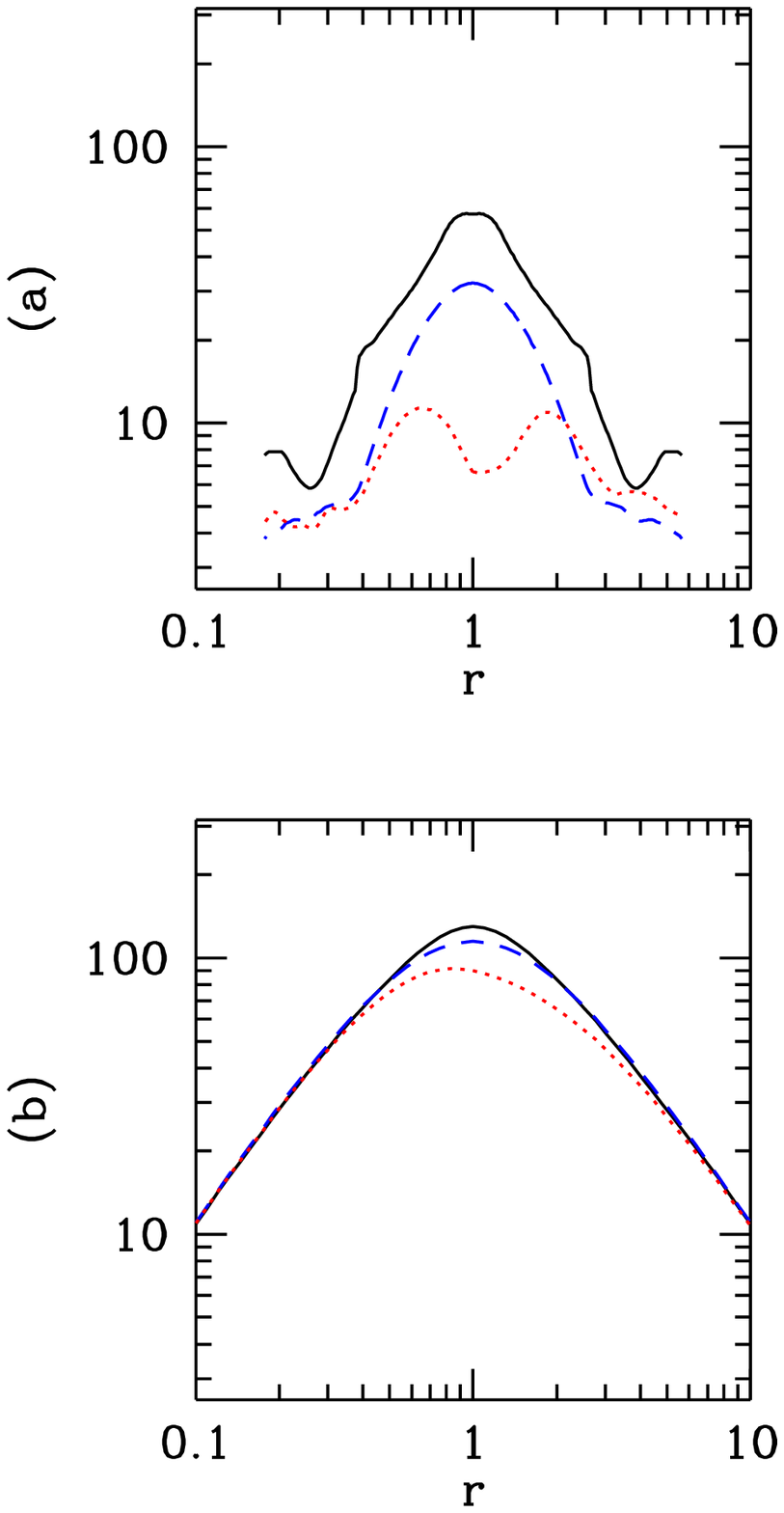,width=5cm}}
\caption{\label{fig5}
The correlators  $C_{ij}(0,r)$ are shown in the same line styles as in
Fig.~\ref{fig4}, but for $z=0$ as function of $r=t'/t$. 
The stronger decoherence of the texture model is even
more evident here.}
\end{figure}

{\em Vector} perturbations are induced by $\bm\Si^{(s)}$ which is
seeded by $\bm{w}^{(v)}$. Transversality and dimensional arguments
require the correlation function to be of the form
\be
 \langle w_i^{(v)}(\bk,t)w_j^{(v)*}(\bk,t')\rangle =
\sqrt{tt'}(k^2\de_{ij}-k_ik_j)W(z,r)~. \label{Cv} 
\ee
Again, as a consequence of causality, the function $W$ is analytic in
$z^2$ (see \cite{DK}). The function $W(z,r)$ is plotted in Fig.~6.
In Figs.~7 and~8 we graph $W(z,1)$ and $W(0,r)$.
\begin{figure}[htb]
\centerline{\psfig{figure=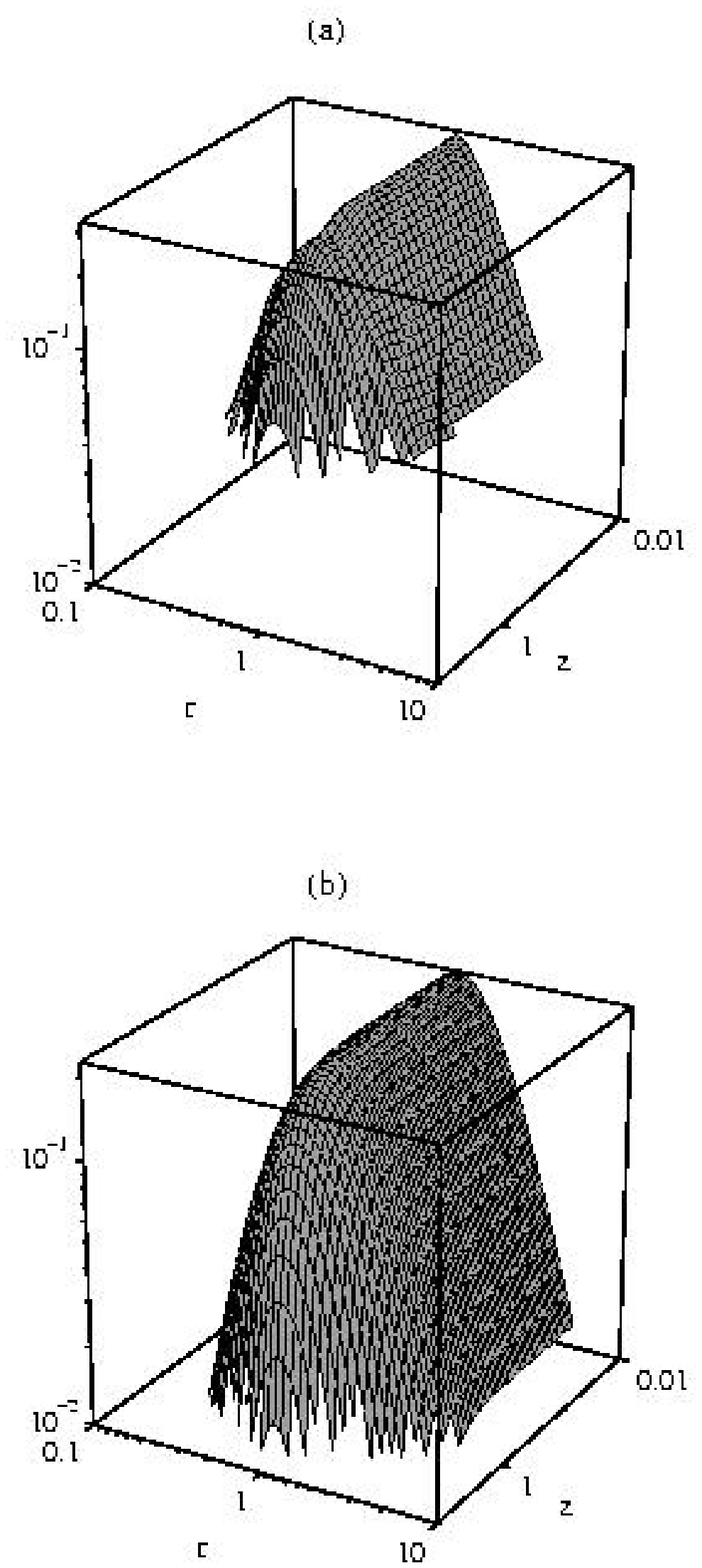,width=5.5cm}}
\caption{\label{fig6}
The vector correlator  $W(z,r)$ is shown. The texture
simulations, panel (a), and the large-$N$ limit, panel (b), give
very similar results.}
\end{figure}
\begin{figure}[htb]
\centerline{\psfig{figure=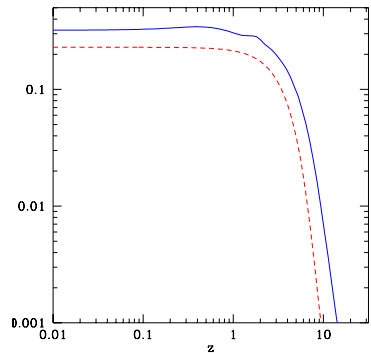,width=5cm}}
\caption{ \label{fig7}
The vector correlator  $W(z,1)$ is plotted. The solid line
represents the texture simulations and the dashed line is the
large-$N$ result. Up to a slight difference in amplitude, the two
results are very similar.}
\end{figure}
\begin{figure}[htb]
\centerline{\psfig{figure=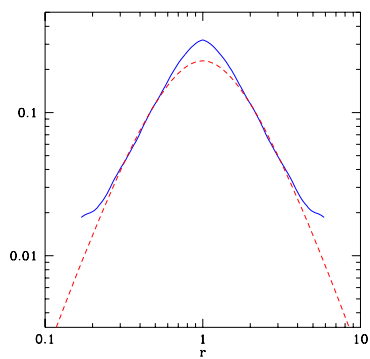,width=5cm}}
\caption{\label{fig8}
The vector correlator  $W(0,r)$ is shown. The solid line
represents the texture simulations and the dashed line is the
large-$N$ result. Also here, the two results are very similar. The
'wings' visible in the texture curve are probably not due to a
resolution problem but the beginning of oscillations.}
\end{figure}

Symmetry, transversality and tracelessness,  together with 
statistical isotropy require the {\em tensor} correlator to be 
of the form (see~\cite{DK})
\bea
\lefteqn{\langle \tau^{(\pi)}_{ij}(t)\tau^{(\pi)*}_{lm}(t')\rangle =} 
\nonumber \\
&&	{1\over\sqrt{tt'}}T(z,r)[\de_{il}\de_{jm}
+\de_{im}\de_{jl}   -\de_{ij}\de_{lm} + 
	k^{-2}(\de_{ij}k_lk_m + \nonumber \\ &&
	\de_{lm}k_ik_j -\de_{il}k_jk_m - \de_{im}k_lk_j -\de_{jl}k_ik_m
	-\de_{jm}k_lk_i) + \nonumber \\
&&	k^{-4}k_ik_jk_lk_m] \label{Ct}
    ~.\eea
The functions $T(z,r)$ as well as $T(z,1)$ and $T(0,r)$ are shown in
Figs.~9 to~11. 
\begin{figure}[htb]
\centerline{\psfig{figure=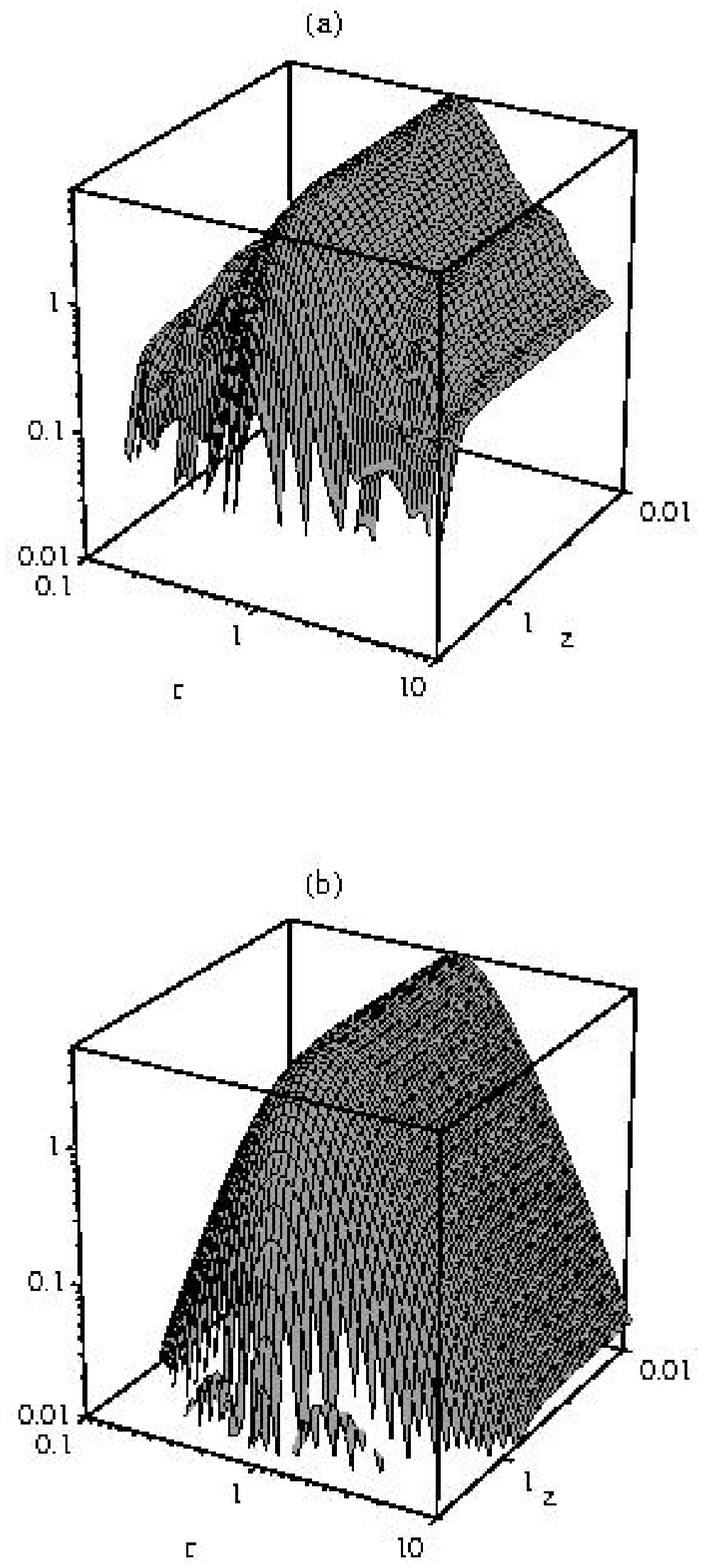,width=5.5cm}}
\caption{\label{fig9}
As Fig.~\ref{fig6}, but for the tensor source function
$T(z,r)$.}
\end{figure}
\begin{figure}[htb]
\centerline{\psfig{figure=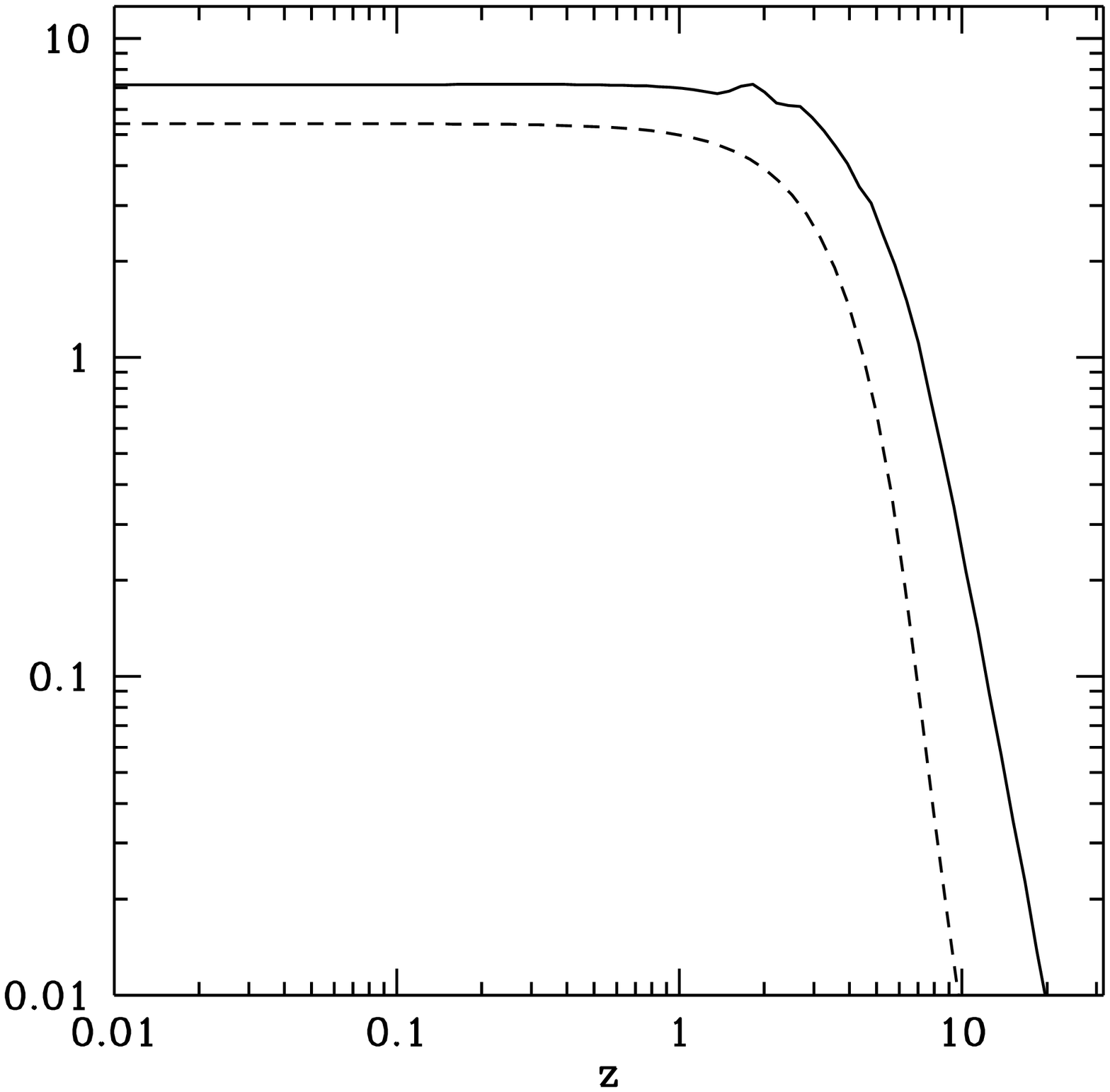,width=5cm}}
\caption{\label{fig10}
As Fig.~\ref{fig7}, but for the tensor source function
$T(z,1)$.}
\end{figure}
\begin{figure}[htb]
\centerline{\psfig{figure=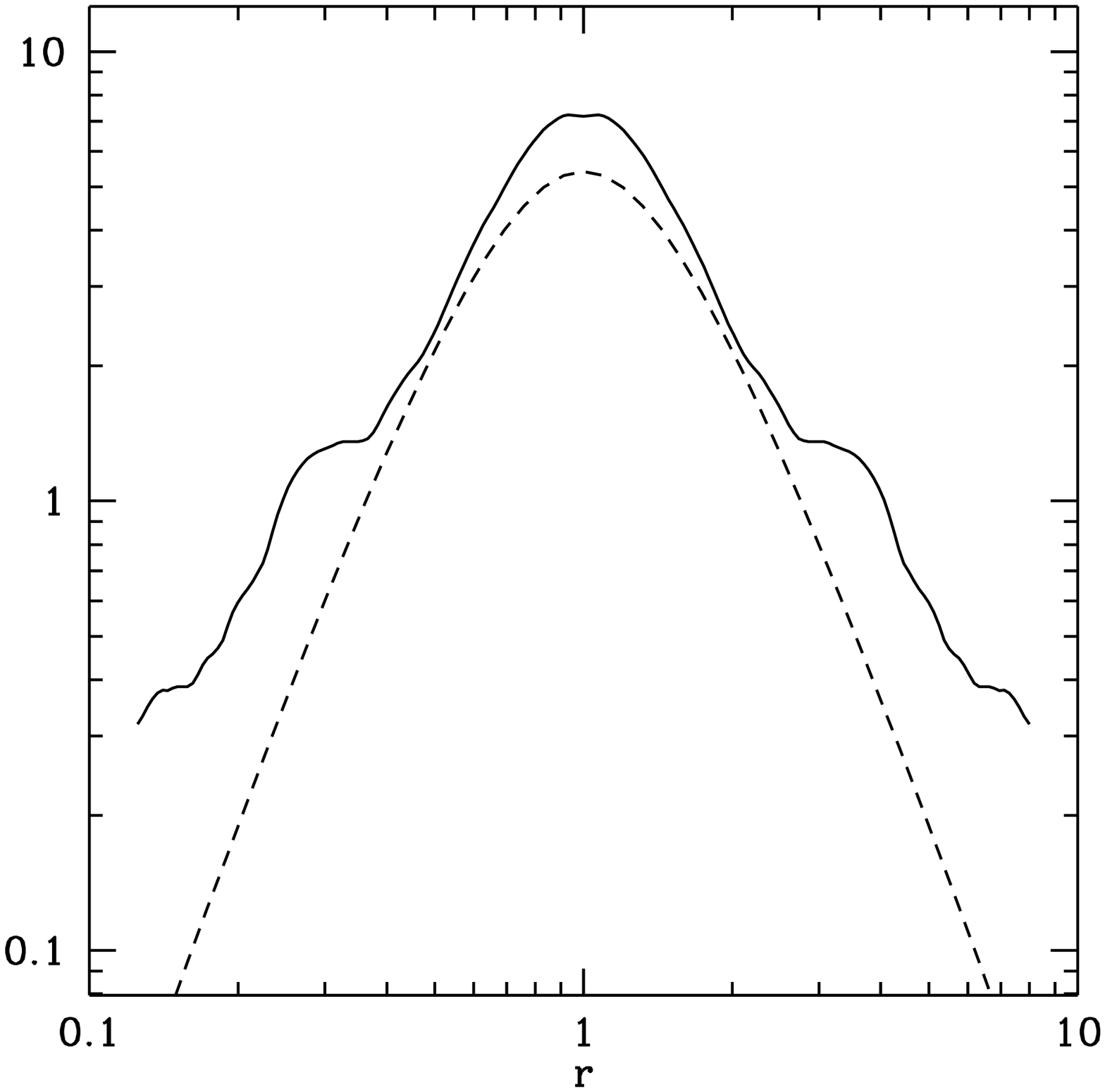,width=5cm}}
\caption{\label{fig11}
As Fig.~\ref{fig8}, but for the tensor source function
$T(0,r)$.}
\end{figure}

Clearly, all correlations between scalar and vector, scalar and tensor
as well as vector and tensor perturbations have to vanish.

\begin{figure}[htb]
\centerline{\psfig{figure=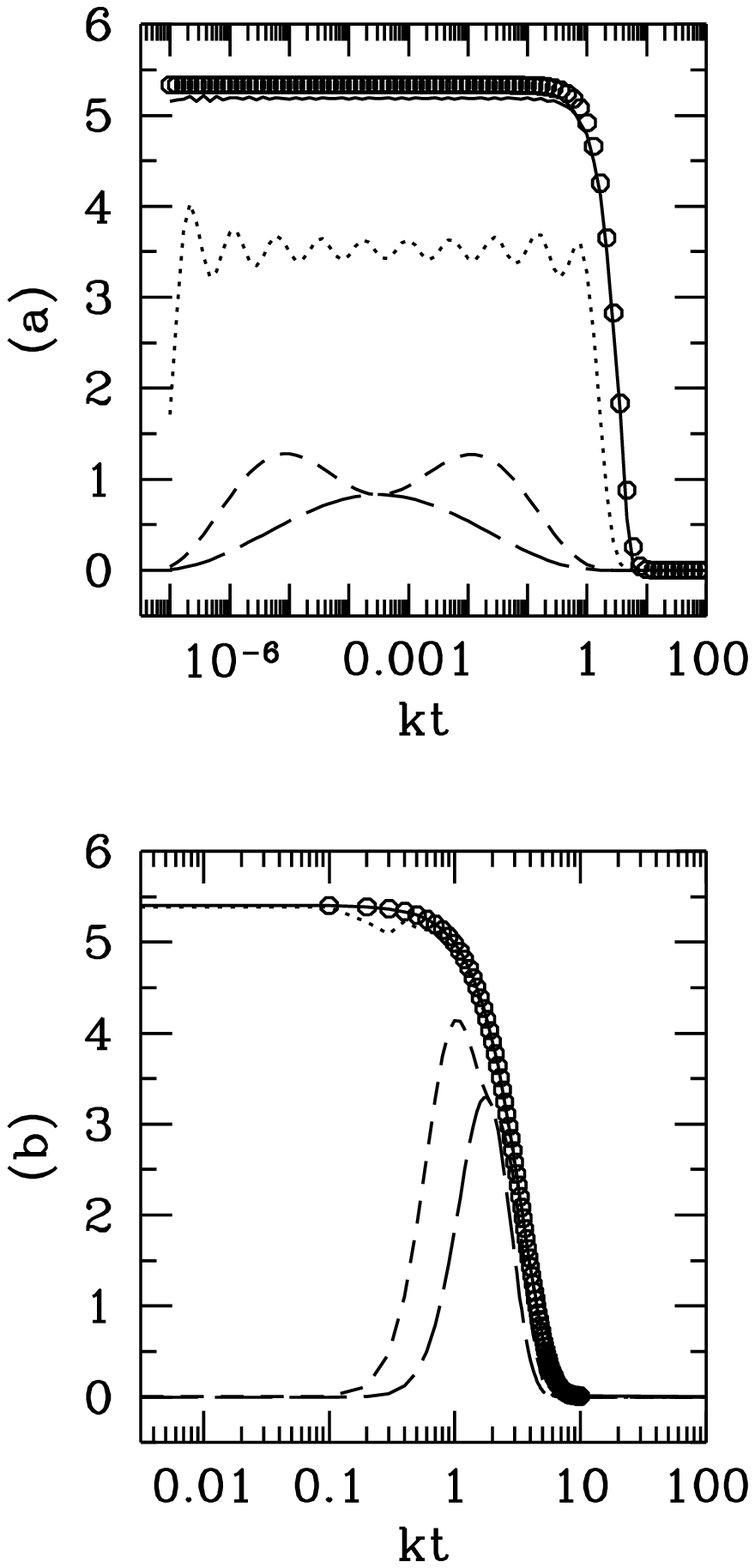,width=5.5cm}}
\caption{\label{fig13}
The sum of the first few eigenfunctions of $T(x,x)$ is shown
for two different weight functions, (a) logarithmic, $w=1/x$ and (b)
linear, $w=1$. The first (long dashed), first and second (short
dashed), first ten (dotted) and first thirty (solid)
eigenfunctions are summed up. The open circles represent the full
correlation function. Clearly, the eigenfunctions obtained by linear  
weighting converge
much faster. Here we only show the equal time diagonal of the
correlation matrix, but the same behavior is also found in the
$C_\ell$ power spectrum which is sensitive to the full correlation matrix.
}
\end{figure}

The scalar source correlation matrix $C$ and the functions $W$ and 
$T$ can be considered as  kernels of positive hermitian operators
in the variables $x=kt=zr^{1/2}$ and $x'=kt'=z/r^{1/2}$, which 
can be diagonalized.
\bea
 C_{ij}(x,x') &=& \sum_n \la_n^{(S)} v_{in}^{(S)}(x)v_{jn}^{(S)\;*}(x')~, 
	\label{EVS}\\   
 W(x,x')  &=& \sum_n \la_n^{(V)} v_n^{(V)}(x)v_n^{(V)\;*}(x')~, 
	\label{EVV}\\
 T(x,x') &=&   \sum_n \la_n^{(T)} v_n^{(T)}(x)v_n^{(T)\;*}(x') 
	\label{EVT} ~,
\eea  
where the series $\left(v_{in}^{(S)}\right)$, $\left(v_n^{(V)}\right)$
and  $\left(v_n^{(T)}\right)$ are 
orthonormal series of eigenvectors (ordered according to the amplitude
of the corresponding eigenvalue) of the operators $C$, $W$ and 
$T$ respectively for a given weight function $w$.
We then have\footnote{Here the assumption that the operators $C$, $W$
and $T$ are trace-class enters. This hypothesis is verified
numerically by the fast convergence of the sums~(\ref{EVS}) to (\ref{EVT}).}
\bea
 \int  C_{ij}(x,x')v_{jn}^{(S)}(x') w(x')dx' &=& \la_n^{(S)}
	v_{in}^{(S)}(x)~, \\ 
 \int  W(x,x') v_n^{(V)}(x')  w(x')dx' &=& \la_n^{(V)}
	v_n^{(V)}(x)~, \\
 \int  T(x,x') v_n^{(T)}(x')  w(x')dx' &=& \la_n^{(T)}
	v_n^{(T)}(x)~.
\eea 
The  eigenvectors  and  eigenvalues depend on the weight
function $w$ which can be chosen to optimize the speed of convergence
of the sums~(\ref{EVS}) to (\ref{EVT}). In our models we found that
scalar perturbations typically need 20 eigenvectors whereas vector
and tensor perturbations need five to ten eigenvectors for an
accuracy of a few percent (see Fig.~\ref{fig13}).

Inserting Eqs.~(\ref{EVS}) to (\ref{EVT}) in Eq.~(\ref{power}), leads
to
\be
\langle X_i(\bk,t_0)X_j^*(\bk,t_0)\rangle = \sum_n\la_n X_i^{(n)}(kt_0)
	X_j^{(n)*}(kt_0) \label{EVpower} ~,
\ee
where $ X_i^{(n)}(t_0)$ is the solution of Eq.~(\ref{lineq}) with
deterministic source term $v_i^{(n)}$.
\be
X_j^{(n)}(t_0,\bk) =\int_{t_{in}}^{t_0}dt\GG(t_0,t,\bk)_{jl}v_l^{(n)}(x,\bk)~.
\label{EVsol}
\ee

For the CMB anisotropy spectrum this gives
\be
 C_\ell = \sum_n^{n_S}\la_n^{(S)}C_\ell^{(Sn)} +  
	\sum_n^{n_V}\la_n^{(V)}C_\ell^{(Vn)}
  +  	\sum_n^{n_T}\la_n^{(T)}C_\ell^{(Tn)} ~.
\ee
$C_\ell^{(\bullet n)}$ is the CMB anisotropy
induced by the deterministic source $v_n$, and $n_{\bullet}$ is the
number of eigenvalues which have to be considered to achieve good accuracy.

Instead of averaging over random solutions of Eq.~(\ref{Gsol}), we can
thus integrate Eq.~(\ref{Gsol}) with the deterministic source term
$v^{(n)}$ and sum up the resulting power spectra. The computational
requirement for the determination of the power spectra of one seed 
model with given source term is thus on the order of
$n_S$ inflationary models.
This eigenvector method has first been applied in Ref.~\cite{PST}.

A source is called totally coherent\cite{Ma,DS} if the unequal time
correlation functions can be factorized. This means that only one 
eigenvector is relevant. A simple totally coherent approximation, 
which however misses some important characteristics of defect
models, can be obtained by replacing the
correlation matrix by the square root
of the product of equal time correlators,
\be \langle \SS_i(t)\SS_j^*(t')\rangle \ra \pm \sqrt{\langle
|\SS_i(t)|^2\rangle  \langle |\SS_j(t')|^2\rangle}  ~. \label{coapp}
\ee 
This approximation is exact if the source evolution is linear. Then the
different $\bk$ modes do not mix and the value of the source term at fixed
$\bk$ at a later time is given by its value at initial time multiplied
by some transfer function, $\SS(\bk,t) =
\SS(\bk,t_{in})T(\bk,t,t_{in})$. In this situation, (\ref{coapp})
becomes an equality and the model is perfectly coherent. Decoherence
is due to the non-linearity of the source evolution which induces a
'sweeping' of power from one scale into another. Different wave
numbers $\bk$ do not evolve independently.

It is interesting to note that the perfectly coherent approximation,
(\ref{coapp}), leaves open a choice of sign which has to be positive
if $i=j$, but which is undetermined otherwise. According to Schwarz
inequality the correlator $ \langle \SS_i(t)\SS_j^*(t')\rangle $ is
bounded by
\bea  \!\!\!\!\!\!
 \lefteqn{- \sqrt{\langle |\SS_i(t)|^2\rangle\langle |\SS_j(t')|^2\rangle}\le
	\langle \SS_i(t)\SS_j^*(t')\rangle \le} \nonumber \\
&&  \sqrt{\langle
	|\SS_i(t)|^2\rangle  \langle |\SS_j(t')|^2\rangle}. \label{bound}
\eea
Therefore, for scales/variables for which the Greens function is not
oscillating (e.g. Sachs Wolfe scales) the full result always lies
between the 'anti-coherent' (minus sign) and the coherent result. We
have verified this behavior numerically.

\begin{figure}[htb]
\centerline{\psfig{figure=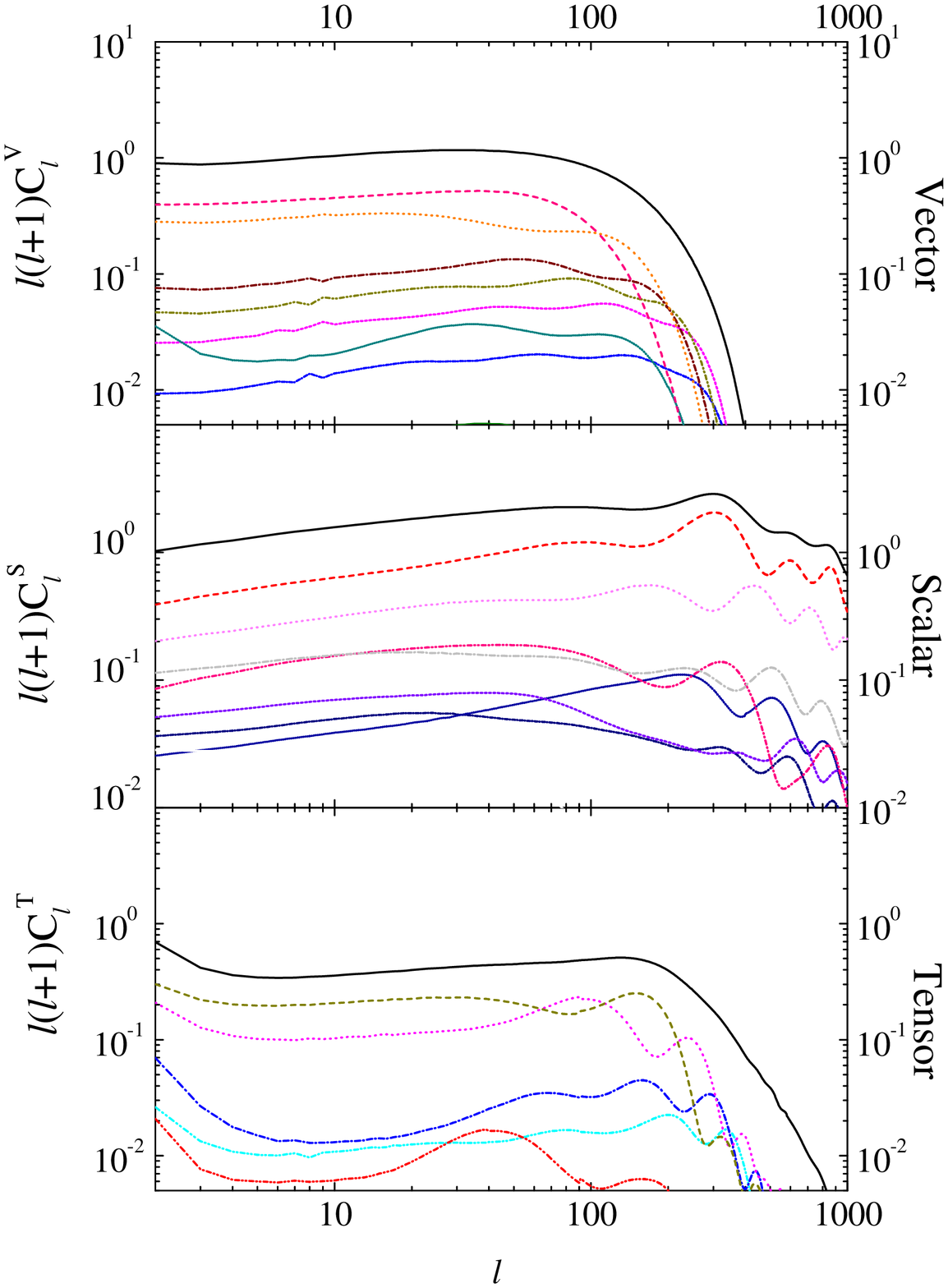,width=6.5cm}}
\caption{\label{fig1416}
The scalar, vector and tensor contributions for the
texture model of structure formation are shown. The dashed lines show
the contributions from single eigenfunctions while the solid line
represents the sum. Note that the single contributions to the scalar and
tensor  spectrum do show oscillations which are however washed out in
the sum. (Vector perturbations do not obey a wave equation and  
thus do not show oscillations.)}
\end{figure}

The first evidence that Doppler peaks are suppressed in defect models
has been obtained in the perfectly coherent approximation in Ref.~\cite{DGS}.
In Fig.~\ref{fig1416} we show the contributions to the $C_\ell$'s from more
and more eigenvectors. A perfectly coherent model has only one
non-zero eigenvalue.

\begin{figure}[htb]
\centerline{\psfig{figure=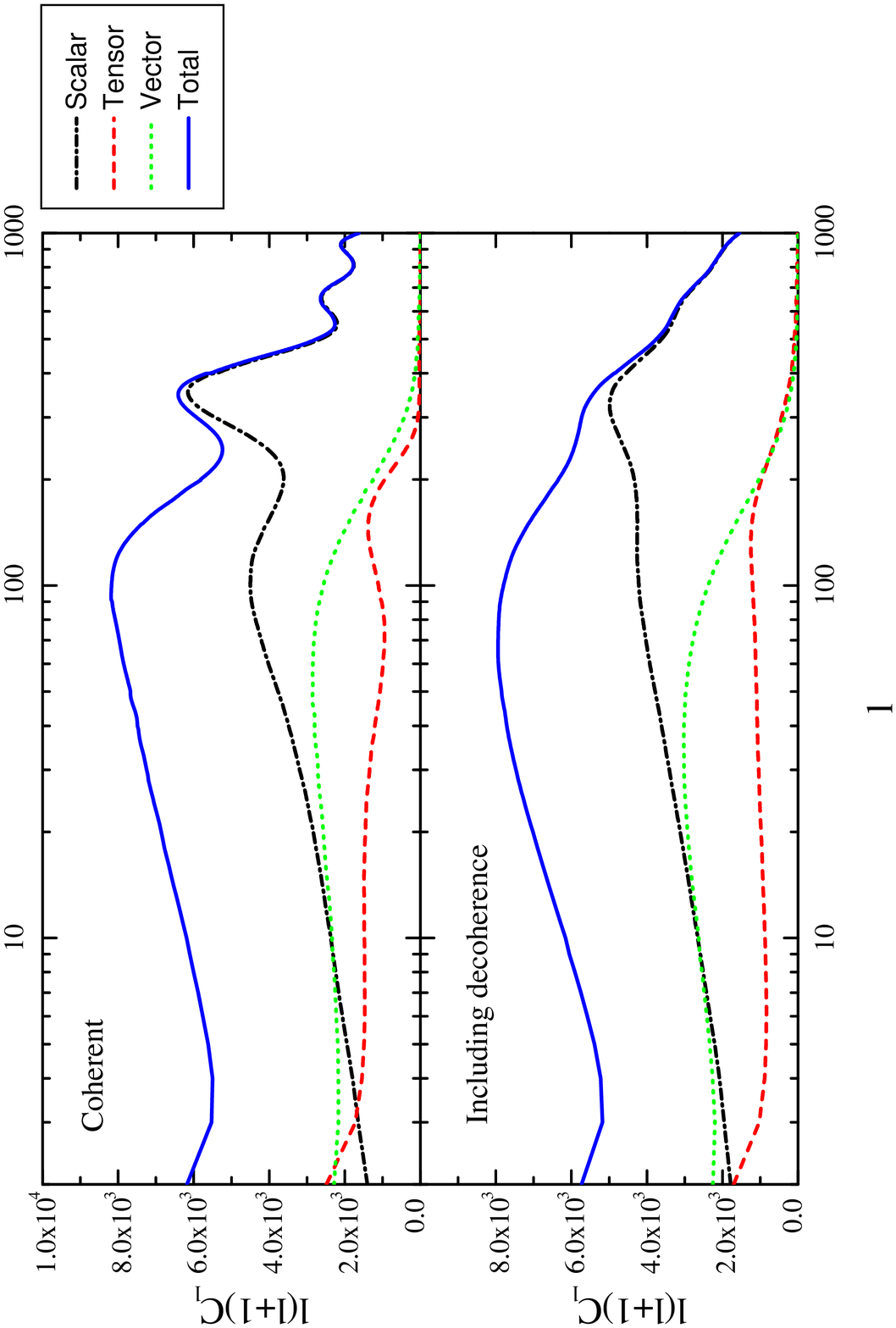,width=5cm,angle=-90}}
\caption{\label{fig17}
The $C_\ell$ power spectrum for the texture scenario is shown
in the perfectly coherent approximation (top panel) and in the full
eigenfunction expansion. Even in the coherent approximation, the
acoustic peaks are not higher than the Sachs Wolfe
plateau. Decoherence just washes out the structure but does not
significantly damp the peaks.}
\end{figure}

A comparison of the full result with the
totally coherent approximation is presented in Fig.~\ref{fig17}. 
There one sees
that decoherence does smear out the oscillations present in the fully
coherent approximation, and does somewhat damp the
amplitude. Decoherence thus prevents the appearance of a series of
acoustic peaks. The absence of power on this angular scale, however,
is not a consequence of decoherence but is mainly due to the
anisotropic stresses of the source which lead to perturbations in the
geometry inducing large scale $C_\ell$'s (Sachs Wolfe), but not to density
fluctuations. Large anisotropic stresses are also at the origin of
vector and tensor fluctuations. 
Our results are in agreement with Refs.~\cite{DGS} and
\cite{PST} but we disagree with Ref.~\cite{TC}, which has found
acoustic peaks with an amplitude of about six in the  coherent
approximation. 

In the real universe, perfect scaling of the seed correlation functions 
is broken by the radiation--matter transition, which
takes place at the time of equal matter and radiation, 
$t_{eq}\simeq20h^{-2}\Om_m^{-1/2}$Mpc. The time $t_{eq}$ is an additional
scale which enters the problem and influences the seed
correlators. Only in a purely radiation or matter dominated universe are
the correlators strictly scale invariant. This means actually that
the $k$ dependence of the correlators $C$, $W$ and $T$ cannot really
be cast into a dependence on $x$ and $x'$, but that these functions depend
on $t,t'$ and $k$ in a more complicated way.
We have to calculate and  diagonalize the seed correlators  
for each wave number $k$ separately and the huge gain of dynamical
range is lost as soon as scaling is lost.

\begin{figure}
\centerline{\psfig{figure=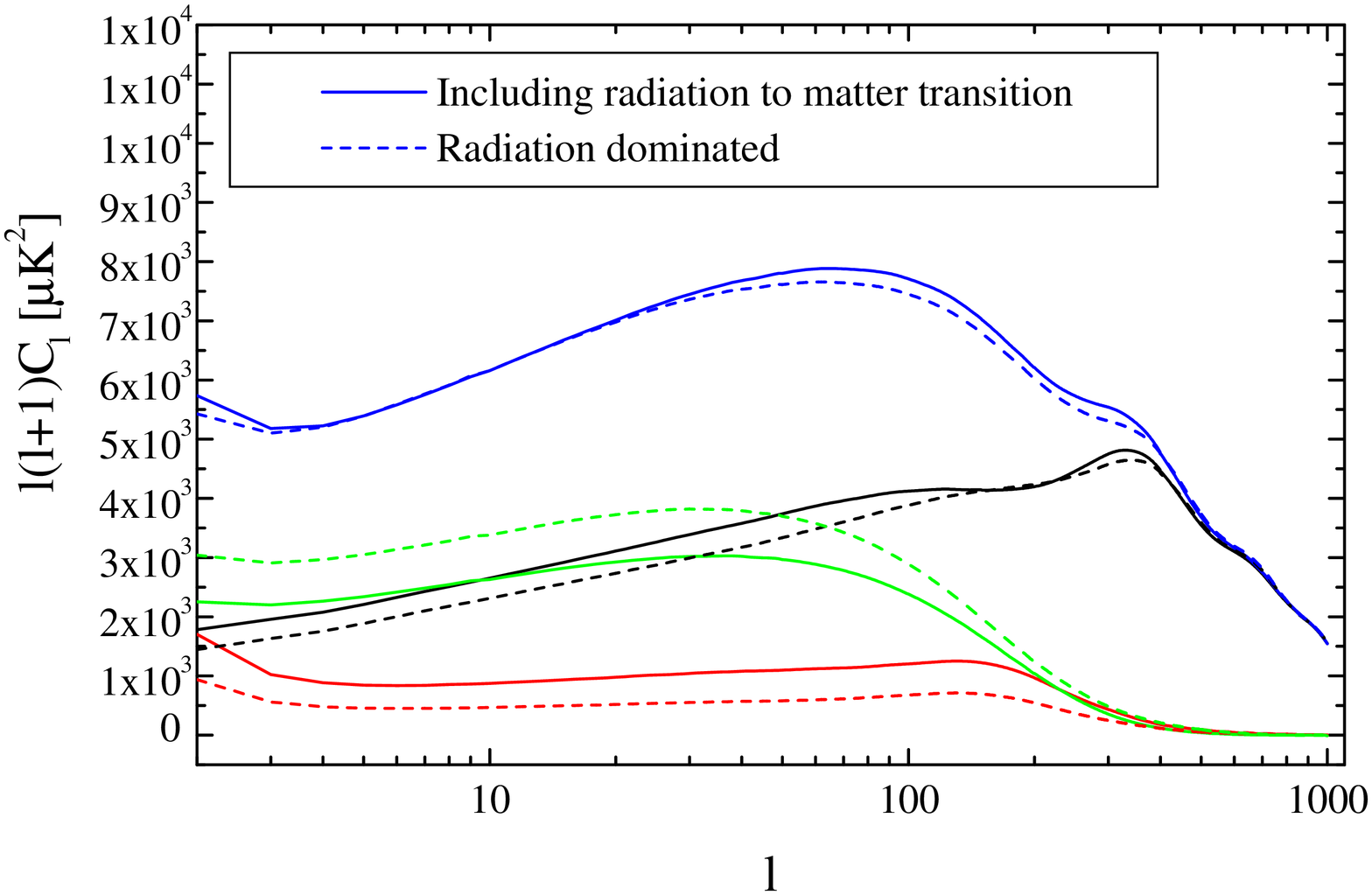,width=5cm,angle=-90}}
\caption{\label{fig18}
The scalar, vector, tensor and total $C_{\ell}$ power spectrum
is shown from pure radiation sources and for an interpolated
source. While the vector perturbations are somewhat higher in the
radiation era, scalar and tensor perturbations are higher in the
matter era and the sum is nearly unchanged.}   
\end{figure}

In the actual case at hand, however, the deviation from scaling is
weak, and most of the scales of interest to us enter the horizon only
in the matter dominated regime. The behavior of the correlators in the
radiation dominated era is of minor importance.
To solve the problem, we calculate the correlator eigenvalues
and eigenfunctions twice, in a pure radiation and in a pure matter
universe and we interpolate the source term from the radiation to the
matter epoch. Denoting by $\la_m, v_m$ and $\la_r, v_r$ a given pair
of eigenvalue and eigenvector in a matter and radiation universe
respectively, we choose as our deterministic source function
\bea
 v(t) &=& y(t)\sqrt{\la_r}v_r(kt) +(1-y(t))\sqrt{\la_m}v_m(kt) \\
\mbox{ with, {\em e.g.},}&& \nonumber\\
 y(t) &=& {t_{eq}\over t+t_{eq}} ~~\mbox{ or }~~~ y(t) =\exp(-t/t_{eq})~,
\eea 
or some other suitable interpolation function.
In Fig.~\ref{fig18} we show the results for scalar, vector and tensor 
perturbations respectively using purely radiation dominated era
and from interpolated source terms. 

Clearly the effect of the radiation dominated early state of the
universe is relatively unimportant for the scales considered here.
The difference between the pure matter era result and the
interpolation is barely visible and thus not shown on the plot. 
This seems to be quite
different for cosmic strings where the fluctuations in the radiation
era are about twice as large as those in the matter era\cite{Paul}.
The radiation dominated era has very little effect on the key results
which we are reporting here; namely the absence of acoustic peaks and
the missing power on very large scales.

In models with cosmological constant, there is actually a second break
of scale invariance at the matter--$\La$ transition. There we
proceed in the same way as outlined above. Since defects cease to
scale and disappear rapidly in an exponentially expanding universe,
the eigenvalues for the $\La$ dominated universe all vanish. 

\subsection{Initial conditions and numerical implementation}

We numerically integrate our system of equations from
redshift $z = 10^7$ up to the present with the goal to have
one percent accuracy up to $\ell \sim 1000$, for a given source term.
 We use the  integration method described in Refs.~\cite{melvit} and 
 \cite{debe}.
We sample the interval $-5 \le \log_{10}kh^{-1}\Mpc \le -0.75$ with
 minimum step size $\Delta \log_{10} = 0.04$, for the scalar case
and  use a smoothing 
algorithm to suppress high frequency sampling noise.
In order to save computing time,
we start the integration of the  $\sigma_{\ell}(k)$'s
with $10$ harmonics, adding new harmonics in the course of the integration.
We find that typically $\sim 40$ harmonics are sufficient for
 small $k$ values ($\log_{10} kh^{-1}\Mpc \lsim -3$), while for higher
$k$ ($\log_{10} kh^{-1}\Mpc \gsim -1$),  up to $\sim 1500$
harmonics for the scalar case, $\sim 200$ for the tensor case are
needed to achieve the desired accuracy.
 Including more than $40$ harmonics for 
 neutrinos corrects our results by less than one percent.
We obtain $\Phi$ algebraically using Eq.~(\ref{Phi}). With this choice of
variables we avoid the numerical difficulties present in
conformal gauge~\cite{maber}, where $\Phi$ is determined by numerical
integration. 

 The abundance of free electrons,
$n_e$, is calculated following a standard recombination scheme
\cite{jones} for $H$ and $^4He$, for a helium abundance by mass of 23\%.
 At high redshift $z \ge 10^5$, the
Thomson opacity is very large, and photons and baryons are tightly coupled.
Due to the large Thomson drag term, Eqs.(\ref{Vb}) 
and (\ref{wb}) become stiff and  difficult to solve numerically. 
Therefore, in this limit
we follow the method of Ref.~\cite{peeyu}, which is accurate to second order
in $(\sigma_T n_e)^{-1}$ (see also \cite{maber}).
Assuming a standard inflationary model, we obtain a single 
scalar power spectrum in few minutes ($\sim 30$ seconds for the tensor case) 
on a PC class workstation, which differs by  less than one percent 
from the $C_\ell$'s computed with other codes \cite{CMBfast},\cite{maber}.

 Summing the scalar $C_\ell$'s from the largest $15$
eigenvectors ($5$ in the tensor case, $10$ for vector perturbations)
typically reproduces the total sum to better than
$5 \%$ (see Fig.~\ref{fig1416}).

\section{The numerical simulations}

As in previous work~\cite{ZD}, we consider a
spontaneously broken scalar
field with O($N$) symmetry. We use the  $\si$-model
approximation, {\em i.e.}, the equation of motion
\be
\Box\b-\l(\b\cd\Box\b\r)\b=0, \label{sigma}
\ee
where $\b$ is the rescaled field $\b=\phi/\eta$.

We do not solve the equation of motion directly, but use a discretized
version of the action~\cite{PST2}:
\be
S=\int d^4\!x \,a^2(t) \l[\frac{1}{2} \dd_\mu \b \cd \dd^\mu \b
	+\frac{\la}{2} \l(\b^2-1\r)\r]\quad,
\ee
where $\la$ is a Lagrange multiplier which fixes the field to the
vacuum manifold (this corresponds to an infinite Higgs
mass). Tests have shown that this formalism agrees
well with the complementary approach of using the equation of
motion of a scalar field with Mexican hat potential and setting
the inverse mass of the particle to the smallest scale that can
be resolved in the simulation (typically of the order of $10^{-35}$ GeV),
but tends to give better energy momentum conservation.

As we cannot trace the field evolution from the unbroken phase
through the phase transition due to the limited dynamical range,
we choose initially a random field at a comoving time $t=2\De x$.
Different grid points are uncorrelated at all earlier
times~\cite{aberna}.

\begin{figure}
\centerline{\psfig{figure=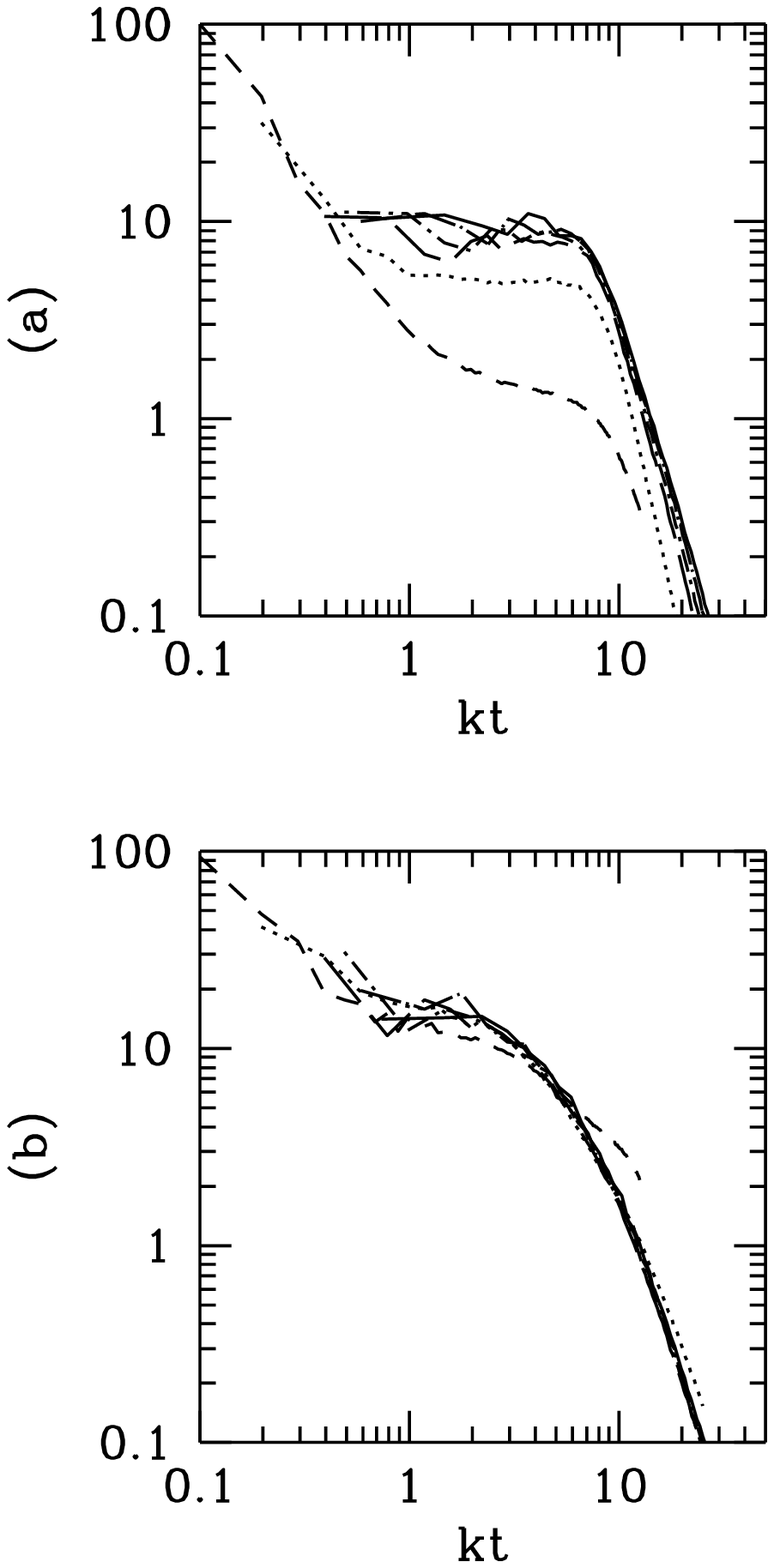,width=5.5cm}}
\caption{\label{fig20}
 The ETCs $C_{11}(z,1)=\langle|\Phi|^2\rangle(kt)$ (panel a) and 
 $C_{22}(z,1)=\langle|\Psi|^2\rangle(kt)$ (panel b) are shown for different
 times. In grid units the times are
 $t=4$ (dashed), $t=8$ (dotted), $t=12$ (long dashed),  $t=16,~20$
 (dash dotted, long dash dotted) and $t=24$ (solid). Clearly $C_{22}$ 
 scales much sooner than $C_{11}$.  To safely arrive in the scaling
 regime one has to wait until $t\sim 16$ and $C_{ij}(kt=0)$ is best
 determined at $t\ge 20$ but $kt< 1$.}
\end{figure}

The use of finite differences in the discretized action as well as in
the calculation of the energy momentum tensor introduce immediately
strong correlations between neighboring grid points. This problem
manifests itself in an initial phase of non-scaling
behaviour, the length of which varies between $10\De x$ and $20\De x$,
depending  on the variable considered.
 It is very important to use
results from the scaling regime only (cf. Fig.~\ref{fig20}).

In order to reduce the time necessary to reach scaling and to
improve the  overall accuracy, we try to 
choose the finite differences in an optimal way. Our current code calculates
all values in the center of each cubic cell defined by the lattice.
The additional smoothing introduced by this  improves energy-momentum
conservation by several percent.\footnote{Julian Borrill suggested 
 to introduce ``spherical derivatives'' that take into account 
the fact that the vacuum manifold is a N-sphere and therefore curved,
 and that this
curvature should be important at least in the initial stages of the
simulation and for unwinding events~\cite{JB}.
So far we haven't investigated this idea sufficiently to include it
into our production code.}

To calculate the unequal time correlator (UTC), the value of the  
observable under consideration is saved once scaling is reached at 
time $t_c$ (we checked this by using different correlation times) and 
then correlated at all following time steps. While there is
some danger of contaminating the equal time correlator (ETC), which contributes
most strongly to the $C_\ell$'s, with non-scaling sources, this method
ensures that the constant for $kt \ra 0$ is determined with maximal
precision for the ETCs.  This is very
important as the constants $C_{ij}(0,1),~W(0,1)$ and $T(0,1)$ fix
the relative size of scalar, vector and tensor contributions of the
Sachs-Wolfe part and severely influence the resulting $C_\ell$'s.
In contrast, the CMB spectrum seems quite stable under small
variations of the shape of the UTCs.

The resulting UTCs are obtained numerically as functions of the 
variables $k$, $t$ and $t_c$ with $t\geq t_c$ and $t_c$ fixed. They 
are then linearly interpolated to the required range. We
construct a hermitian $100\times 100$ matrix in $kt$ and $kt'$, 
with the values of $kt$ chosen on a linear scale to maximize the 
information content, $0\leq kt\leq x_{\mathrm max}$. The choice of 
a linear scale ensures good convergence
of the sum of the eigenvectors after diagonalization 
(see Fig.~\ref{fig13}), but still retains
enough data points in the critical region, ${\cal O}(x) =1$,
 where the correlators
start to decay. In practice we choose as the endpoint $x_{\mathrm max}$
of the range sampled by the simulation the value at which the correlator
decays by about two orders of magnitude, typically $x_{\mathrm max}\approx 40$.
The eigenvectors that are fed into the Boltzmann code
are then interpolated using cubic splines with the condition
$v_n(kt)\ra 0$ for $kt \gg x_{\mathrm max}$.

\begin{figure}
\centerline{\psfig{figure=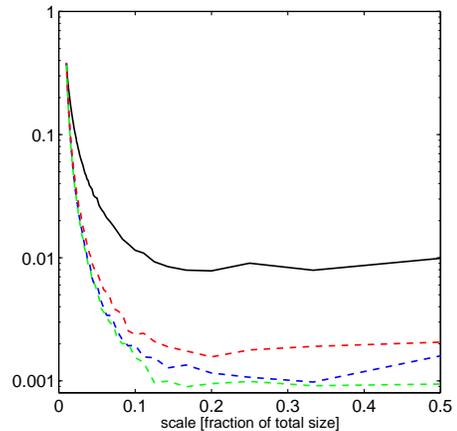,width=6.5cm}}
\caption{\label{fig21}
Energy momentum conservation of our numerical simulations
is shown. The lines represent the sum of the terms which has to vanish
if energy (solid) respectively momentum (dashed) is conserved,
divided by the sum of the absolute value of these terms. The abscissa
indicates the wavelength of the perturbation as fraction of the size
of the entire grid. }
\end{figure}

We use several methods to test the accuracy of the simulation: 
energy momentum conservation of the defects code is found to be
better than 10\% on all scales larger than about 4 grid units, as is 
seen in Fig.~\ref{fig21}. A comparison with the exact
spherically symmetric solution in non-expanding space \cite{PST2}
shows very good agreement. 

\begin{figure}
\centerline{\psfig{figure=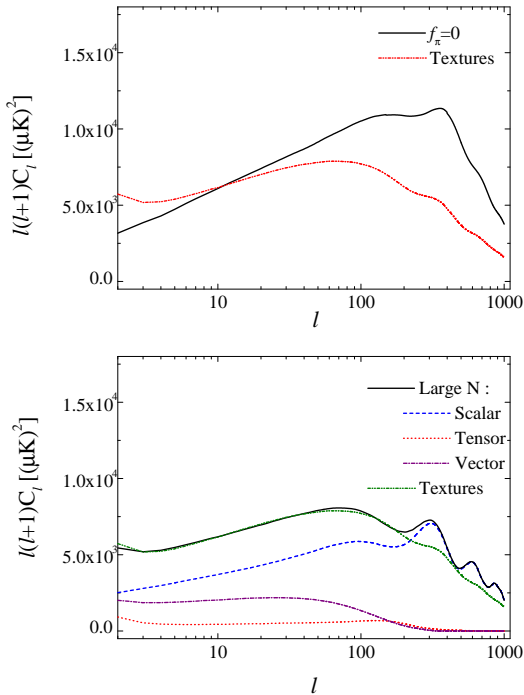,width=7cm}}
\caption{\label{fig19}
Top panel: the ${\it f_{\pi}}=0$ model.
Bottom panel: The $C_\ell$ power spectrum is shown for the large-$N$ limit
(bold line) and for the texture model. The main difference is
clearly that the large-$N$ curve shows some acoustic oscillations
which are nearly entirely washed out in the texture case.}
\end{figure}

The resulting CMB spectrum on Sachs Wolfe scales is  consistent with 
the line of sight integration of Ref.~\cite{ZD}.
Furthermore, the overall shape and amplitude of the unequal time correlators
are quite similar to those found in the analytic large-$N$ approximation
\cite{TS,KD,DK}  (see Figs.~1 to 11). The main difference of the 
large-$N$ approximation is
that there the field evolution, Eq.~(\ref{sigma}), is approximated by a
linear equation. The non-linearities in the large-$N$ seeds which are
due solely to the energy momentum tensor being quadratic in the
fields, are much weaker than in the texture model where the field
evolution itself is non-linear. Therefore, decoherence which is a
purely non-linear effect, is expected to be much weaker in the
large-$N$ limit. This is actually the main difference between the two
models as can be seen in Fig.~\ref{fig19}.

\section{ Results and comparison with data}

\subsection{CMB anisotropies}

The $C_\ell$'s for the 'standard' global texture model are shown in 
Fig.~\ref{fig17}, bottom panel.
\begin{figure}
\centerline{\psfig{figure=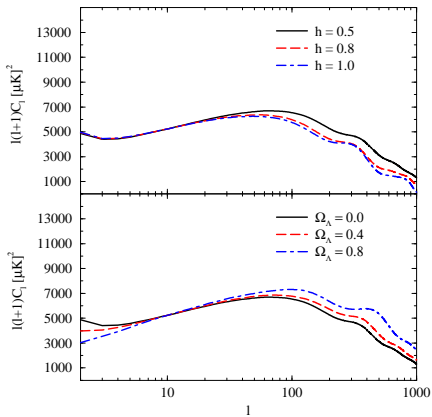,width=7cm}}
\caption{\label{fig23}
The $C_\ell$ power spectrum is shown for different values of
cosmological parameters. In the top panel we choose $\Om_\La=0$,
$\Om_{CDM}=0.95$, $\Om_b=0.05$ and vary $h$. In the bottom panel we
fix $h=0.5$, $\Om_b=0.05$ and vary  $\Om_\La$. We only consider
spatially flat universes, $\Om_0=1$.}
\end{figure}

\begin{table}[ht]
\begin{center}
\begin{tabular}{||c|c|c|c||}
\hline
$\Om_\La$ & $h$ & $\ep$ & $\si_8$\\
\hline
 0.0 & 0.5 & $(1.66 \pm 0.17) 10^{-5}$ & 0.24 \\
 0.0 & 0.8 & $(1.67 \pm 0.17) 10^{-5}$ & 0.34 \\
 0.0 & 1.0 & $(1.68 \pm 0.17) 10^{-5}$ & 0.44 \\
 0.4 & 0.5 & $(1.64 \pm 0.16) 10^{-5}$ & 0.22 \\
 0.8 & 0.5 & $(1.59 \pm 0.16) 10^{-5}$ & 0.16  \\
\hline
\end{tabular}
\end{center}
\caption{\label{tabeps}
The value of the normalization constant $\ep$ and the fluctuation
amplitude $\si_8$ are given for
the different models considered. The error in $\ep$ comes
from a best fit normalization to the full CMB data set.
 Cosmological parameters which are not
indicated are identical in all models or given by
$\Om_0=\Om_{cdm}+\Om_\La+\Om_b=1$. We consider only spatially flat
models with $\Om_b=0.05$ and a helium fraction of 23\%. The parameter
choice indicated in the top line is referred to as {\em standard}
texture model in the text.}
\end{table} 

Vector and tensor modes are found to be of the same order as 
the scalar component at COBE-scales.
 For the 'standard' texture model 
we obtain $C_{10}^{(S)} : C_{10}^{(V)} : C_{10}^{(T)} \sim  0.9 : 1.0 : 0.3$,
in good agreement with the predictions of Refs.~\cite{Aetal}, 
\cite{PST,Aetal,ABR} and \cite{DK}.
Due to tensor and vector contributions, even assuming perfect coherence 
(see Fig.~\ref{fig17}, top panel), the total power spectrum does not  
increase from large to small scales.
Decoherence leads to smoothing of  oscillations in the power
spectrum at small scales and the final power spectrum has a smooth
shape with a broad, low isocurvature 'hump' at $\ell\sim 100$ and a
small residual of the first acoustic peak at $\ell \sim 350$.
There is no structure of peaks at small scales. The power
spectrum is well fitted by the following fourth-order 
polynomial in $x= \log \ell$:

\be {\ell ( \ell +1 ) C_{\ell}\over 110C_{10}} = 1.5 -2.6 x +
3.3 x^2 -1.4 x^3 +0.17x^4 ~.\ee

The effect of decoherence is less important
for the large-$N$ model, where oscillations and peaks are still visible
 (see Fig~\ref{fig19}, bottom panel). This is due to the fact that 
the non-linearity of the large-$N$ limit is  only in the quadratic 
energy momentum tensor. The scalar field evolution is linear in 
this limit\cite{TS}, in contrast to
the $N=4$ texture model. Since decoherence is inherently due to
non-linearities, we expect it to be stronger for lower values of $N$. 
COBE normalization leads to $\epsilon = (0.92 \pm 0.1) 10^{-5}$
 for the large-$N$ limit.

In Fig.~\ref{fig23} we plot the global texture $C_{\ell}$ power spectrum 
for different choices of cosmological parameters. The  variation of 
parameters leads to  similar effects like in the inflationary case,
but with  smaller amplitude.
At small scales ($\ell \ge 200$), the $C_\ell$'s tend to decrease with
increasing $H_0$ and they increase when a cosmological
constant $\Omega_{\Lambda} = 1 - \Omega_m$ is introduced.
Nonetheless, the amplitude of the anisotropy power spectrum at
high $\ell$s remains in all cases on the same level like the one at
low  $\ell$s, without showing the substantial peak found in 
inflationary models.
The absence of acoustic peaks is a stable prediction
of global $O(N)$ models.
 The models are normalized to the full CMB data set,
 which leads to slightly larger values of the normalization 
parameter $\ep=4\pi G\eta^2$ than pure COBE normalization.
In Table~\ref{tabeps} we give the cosmological parameters and the value of
$\ep$ for the models shown in Fig.~\ref{fig23}.

\begin{figure}
\centerline{\psfig{figure=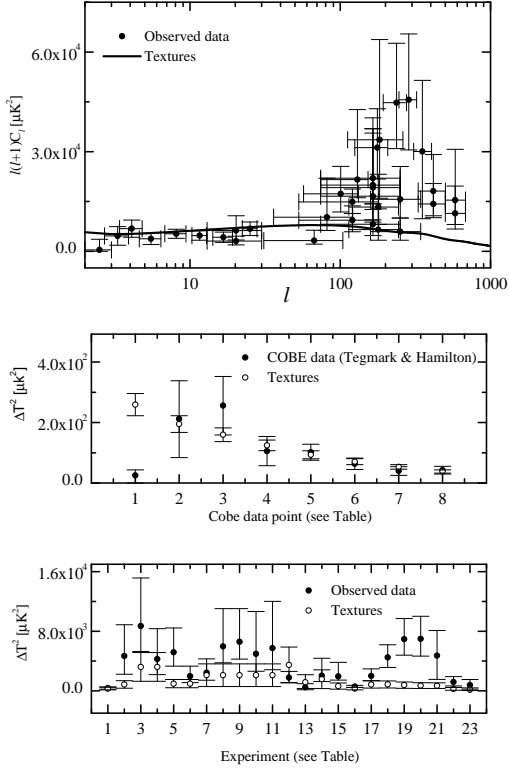,width=7.5cm}}
\caption{\label{fig24}
The $C_\ell$ spectrum obtained in the standard texture model
is compared with data. In the top panel experimental results and the
theoretical curve are shown as functions of $\ell$. In the two lower
panels we indicate the value of each of the 31 experimental data
points with 1-$\si$ error bars  and the  corresponding theoretical value
with its uncertainty. The experiments corresponding to a given number are
given in Table~\ref{tabdata}. In the middle panel the 8 COBE data
points are shown. In the bottom panel other experiments are presented.}
\end{figure}

In order to compare our results with current experimental data, we have
selected a set of $31$ different anisotropy
detections obtained by different experiments, or by the same
experiment with different window functions and/or at different
frequencies. Theoretical predictions and data of CMB anisotropies are 
usually compared by plotting the 
theoretical $C_{\ell}$ curve along with the CMB measurements
converted to band power estimates. We do this in the top panel
of Fig.~\ref{fig24}. The data points  show
an increase in the anisotropies from large to smaller scales, in 
contrast to the theoretical predictions of the model.
This fashion of presenting the data is surely correct, but
lacks informations about the uncertainties in the theoretical model.
Therefore we also compare the detected mean square anisotropy,
 $\Delta^{(Exp)}$ and the experimental 1-$\sigma$ error,
$\Sigma^{(Exp)2}$,  directly 
with the corresponding theoretical mean square anisotropy, given by
\be 
\Delta^{(Th)} = {1 \over {4 \pi}} \sum_{\ell} (2 \ell +1) C_{\ell}
W_{\ell}~,
\ee
where the window function $W_{\ell}$ contains all experimental details
(chop, modulation, beam, etc.) of the experiment.

The theoretical error in principle depends on the statistics of the 
perturbations. If the  distribution is Gaussian, one can associate a 
sample/cosmic variance
\be \Sigma^{(Th)2}={1 \over f}{1\over 8 \pi^2} \sum_{\ell}
(2 \ell +1)W^2_{\ell}C^2_{\ell}~,
  \label{Gauss}\ee
where $f$ represents the fraction of the sky sampled by a given experiment.

 Deviation from Gaussianity leads to an enhancement of this
variance, which can be as large as a factor of $7$ (see \cite{xiao}).
Even if the perturbations are close to Gaussian (which has
been found by simulations on large scales~\cite{ZD,ACSSV}), 
the $C_\ell$'s, which are the squares of Gaussian variables, are 
non-Gaussian. This effect is, however only relevant for relatively low
$\ell$s.
Keeping this caveat in mind, and missing a more precise alternative, we
nevertheless indicate the minimal, Gaussian error calculated
according to~(\ref{Gauss}).
We add a $30 \%$ error from the CMB normalization. The numerical
seeds are assumed to be about 10\% accurate. 

In Table~\ref{tabdata}, the detected mean square anisotropy,
 $\Delta^{(Exp)}$, with the experimental 1-$\sigma$ error 
 are listed for each experiment of our data set. The
 corresponding sky coverage is also indicated.
 In Fig.~\ref{fig24} we plot these data points, together with the 
 theoretical predictions for a texture model with $h=0.5$ and 
 $\Omega_{\Lambda} = 0$.
 
We find that, apart from the
COBE quadrupole, only the Saskatoon experiment disagrees
significantly, more than $1\si$, with our model. But also this
disagreement is below $3\si$ and thus not sufficient to rule
out the model. 
In the last column of Table~\ref{tabdata} we indicate
\[ \chi_j^2= (\De_j^{(Th)}-\De_j^{(Exp)})^2/(\Si_j^{(Th)2}+\Si_j^{(Exp)2}) \]
for the $j$-th experiment, where the theoretical model is the standard
texture model with
$\Omega_{\Lambda}=0$ and $h=0.5$. The major discrepancy
between data and theory comes from the COBE quadrupole.
Leaving away the quadrupole, which can be contaminated 
and leads to a similar $\chi^2$ also for  inflationary models, 
the data agrees quite well with the model, with the exception of 
three Saskatoon  data points.
Making a rough chi-square analysis, we obtain
(excluding the quadrupole) a value $\chi^2=\sum_j\chi_j^2 \sim 30$ for
a total of 30 data points and one constraint. An absolutely reasonable
value, but one should take into account that
the experimental data points which we are considering
are not fully independent. The regions of sky sampled by the
Saskatoon and MSAM  or COBE and Tenerife,
 for instance, overlap.
Nonetheless, even reducing the degrees of freedom of our analysis
to $N = 25$, our $\chi ^2$ is still in the range $(N-1) \pm {
\sqrt 2 (N-1)} \sim 24 \pm 7$ and hence still compatible with 
the data.

This shows that even assuming Gaussian statistics,
 the models are not convincingly ruled out from present CMB data. 
There is however one caveat in this analysis: A chi-square test is not
sensitive to the sign of the discrepancy between theory and
experiment. For our models the theoretical curve is systematically 
lower than the experiments. For example, whenever the discrepancy between
theory and data is larger than $0.5\si$, which happens with nearly half
of the data points (13), in all cases except for the COBE quadrupole, 
the theoretical value is smaller than the data. If
smaller and larger are equally likely, the probability to have $12$ or
 more equal signs  is $2(13+1)/2^{13}\simeq 3.4\times 10^{-3}$. This  
indicates that either the model
is too  low or that the data points are systematically too high. The
number  $0.003$ can however not be taken seriously, because we can
easily change it by  increasing our normalization on a moderate cost of
$\chi^2$.

\subsection{Matter distribution}

In Table~\ref{tabeps} we show the expected variance 
of the total mass 
fluctuation $\sigma_R$ in a ball of radius $R=8h^{-1}$Mpc,
for different choices of cosmological parameters.
We find $\sigma_8 = (0.44 \pm 0.07)h$
(the error coming from the CMB normalization) for a flat
model without cosmological constant, in agreement
with the results of Ref.~\cite{PST}.
From the observed  cluster abundance, one infers 
$\sigma_8 = (0.50 \pm 0.04) \Omega^{-0.5}$ \cite{eke} and
$\sigma_8 = 0.59^{+0.21}_{-0.16}$ \cite{lid}.  These results,
 which are obtained with the Press-Schechter formula, assume Gaussian 
 statistics. We thus have to take them with a
grain of salt, since we do not know how non-Gaussian fluctuations on
cluster scales are in the texture model. According to Ref.~\cite{free}, the 
Hubble constant lies in the interval
$h\simeq 0.73 \pm 0.06 \pm 0.08$. Hence,
in a flat CDM cosmology, taking into account the  uncertainty of
the Hubble constant, the texture scenario predicts a reasonably consistent 
value of $\sigma_8$.

As already noticed in Refs.~\cite{ABR} and \cite{PST},
unbiased global texture models are unable to reproduce
the power of galaxy clustering at very large scales,
 $\gsim 20 h^{-1}$ Mpc.
In order to quantify this discrepancy we compare our prediction of the linear
matter power spectrum with the results from a number of infrared 
(\cite{fish},\cite{tadr}) and optically-selected (\cite{daco}, \cite{lin})
galaxy redshift surveys, and with the real-space power
spectrum inferred from the APM photometric sample (\cite{baugh}) (see 
Fig.~\ref{fig26}). Here, cosmological parameters have important
effects on the shape and amplitude of the matter power spectrum.
Increasing the Hubble constant shifts the peak of the power spectrum 
to smaller scales (in units of $h/$Mpc), while the inclusion of a cosmological
constant enhances large scale power. 

 We consider a set of models in 
 $\Omega_{\Lambda}$ -- $h$ space, with linear bias \cite{kais} as
additional parameter. In  Table~\ref{tabpow} we report
  for each survey and for each model the best value of the bias
parameter obtained by $\chi^2$-minimization. We also indicate the
value of $\chi^2$ (not divided by the number of data points).
The data points and the theoretical predictions are plotted in 
Fig.~21.
Our bias parameter  
strongly depends on the data considered. This is not surprising, since
also  the catalogs are biased relative to each other.

Models without cosmological constant and with $h \sim 0.8$ only require
a relatively modest bias $b \sim 1.3 -3$. But for these models the
shape of the power spectrum is wrong as can be seen from the
value of  $\chi^2$ which is much too large.
The bias factor is in agreement with our prediction for $\sigma_8$.
 For example, our best fit for the IRAS data, for 
 $h \sim 0.8$ is $b \sim 1.3$. With $\sigma_8^{IRAS} = 
(0.69 \pm 0.05)$, this gives  $\sigma_8 \sim 0.48 \pm 0.04$,
compatible with the direct computation

Whether IRAS galaxies are biased  is still under debate. Published values for
the $\beta$ parameter, defined as $\beta= \Omega^{0.6}/b$,
for IRAS galaxies, range between
$\beta_I = 0.9^{+0.2}_{-0.15}$ \cite{kais2} and $\beta_I = 0.5 \pm
0.1$ \cite{will}.  Biasing of IRAS galaxies is
also suggested by measurements of bias in the
optical band. For example, Ref.~\cite{pea97} finds
 $\beta_o =0.40 \pm 0.12$, in marginal agreement 
 with \cite{shai}, which obtains $\beta_o = 0.35 \pm 0.1$.
A bias for IRAS galaxies is not only possible but
even preferred in {\em flat} global texture models. 

But also with  bias, our models are in significant contradiction
with the shape of the power spectrum at large scales.
As the values of $\chi^2$ in Table~\ref{tabpow} and Fig.~\ref{fig26}
 clearly indicate, the models are inconsistent with the shape of the IRAS
power spectrum, and they can be rejected with a high confidence
 level. The APM data which  has the smallest error bars is the most
 stringent evidence against texture models. Nonetheless, these data points
 are not measured in redshift space but they 
 come from a de-projection of a $2-D$ catalog into $3-D$ space.
 This might introduce systematic errors and thus the errors of APM may
be underestimated.
  
Models with a cosmological constant agree much better with the shape
 of the observed power spectra, the value of  $\chi^2$ being low for
 all except the APM data.
But the  values of the bias factors are extremely high for these
 models.  For example, IRAS galaxies should have a bias $b \sim 3 -
 6$,   resulting in $\sigma_8 \le 0.25$, and in a $\beta_{I} \le 0.2$
 which is  too small, even allowing for big variances due to
 non-Gaussian statistics.

The power spectra for the large-$N$ limit and for the coherent
approximation are typically a factor 2 to 3 higher (see Fig.~22), and
the biasing problem is  alleviated for these cases. For
$\Om_\La=0$ we find $\si_8=0.57h $ for the
large-$N$ limit and  $\si_8=0.94h$ for the coherent approximation.
This is no surprise since only one source function, $\Psi_s$, the
analog of the Newtonian potential, seeds dark matter fluctuations and
thus the coherence always enhances the unequal time correlator. The second
inequality in~(\ref{bound}) applies. The dark matter Greens function
is not oscillating, so this enhancement translates directly into
the power spectrum. 

\begin{figure}
\centerline{\psfig{figure=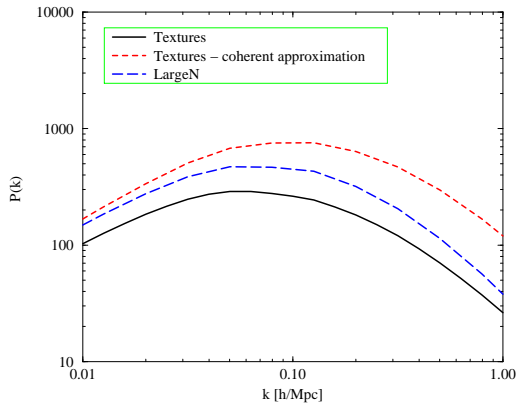,width=7.5cm}}
\caption{\label{fig22}
The dark matter power spectrum for the texture model (solid line) is
compared with the coherent approximation (short dashed) and the
large-$N$ limit (long dashed). The spectra are COBE normalized and the
cosmological parameters are $\Om_\La=0~,~h=0.5$.}
\end{figure}

Models which are anti-coherent in the sense defined in Section~IID
reduce power on Sachs-Wolfe scales and enhance the power in the dark
matter. Anti-coherent scaling seeds are thus the most promising
candidates which may cure some of the problems of global $O(N)$ models.

The simple analysis carried out here does not take into account
the effects of non-linearities and redshift distortions.
Redshift distortions in the texture case should be less important
than in the inflationary case since the peculiar velocities are rather
low (see next paragraph).
 Non-linearities typically set in at $k \ge 0.5h$Mpc$^{-1}$ and should not
have a big effect on our main conclusions which come from much larger
scales.  Inclusion of these corrections will
result in more small-scale power and in a broadening of the spectra,
which even enhances the conflict between models and data.
Furthermore, variations of other cosmological parameters, like
the addition of massive neutrinos, hot dark matter, which is not 
considered here,
will result in a change of the spectrum on small scales but will not
resolve the  discrepancy at large scales.

Nonetheless, scale dependent biasing may exist and 
lead to a non-trivial relation between the calculated dark matter 
power spectrum and the observed galaxy power
spectrum. We are thus very reluctant to rule out the model
by comparing two in principle different things, the relation of which
is far from understood. Therefore we would prefer to reject the models
on the basis of peculiar velocity data, which is more difficult to
measure but most certainly not biased. 

\begin{figure}
\centerline{\psfig{figure=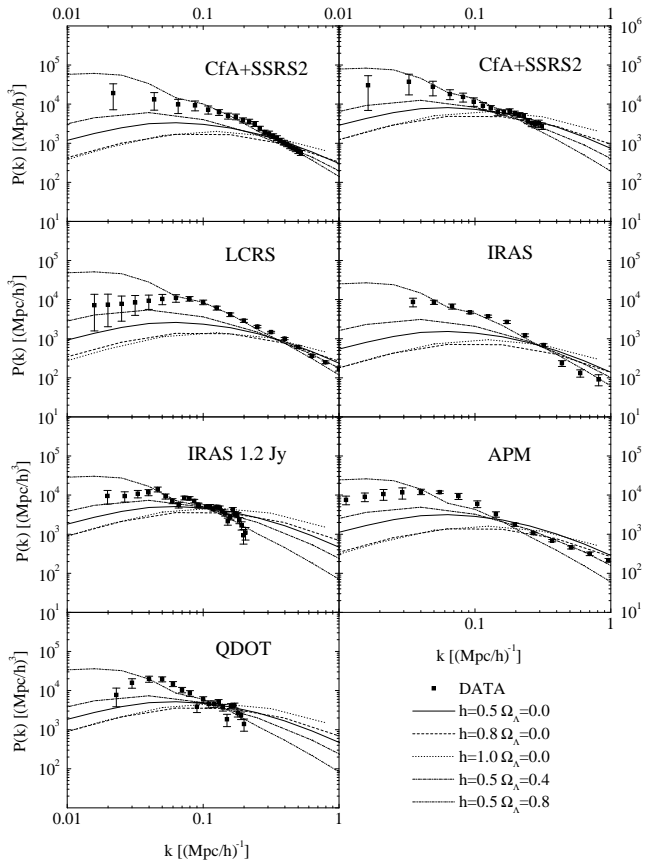,width=8.5cm}}
\caption{\label{fig26}
Matter Power spectrum: comparison between data and theory.
References are in the text. Data set courtesy of M. S. Vogeley 
\protect\cite{voegel}.}
\end{figure}

\subsection{Bulk velocities}

To get a better handle on the missing power on 20 to 100$h^{-1}$Mpc,
 we investigate the
velocity power spectrum which is not plagued by biasing problems. The
assumption that galaxies are fair tracers of the velocity field seems
to us much better justified, than to assume that they are fair tracers
of the mass density. 
We therefore test our models against peculiar velocity data. We
use the data by Ref.~\cite{deke} which gives the bulk flow

\be \si^2_v(R) = {H_0^2\Om_m^{1.2} \over {2 \pi^2}} \int P(k)W(kR)dk ~,\ee

in spheres of radii $R = 10$ to $60 h^{-1}$Mpc. 
These data are derived after reconstructing the $3-$dimensional
velocity field with the POTENT method (see~\cite{deke} and references 
therein).

As we can see from Table~\ref{tabevl}, the COBE normalized texture 
model predicts too low velocities on large scales when compared with
POTENT results.
Recent measurements of the bulk flow lead to somewhat lower estimates
like $\sigma_{v}(R) \sim (230 \pm 90)$ at $R = 60 h^{-1}$Mpc
(\cite{giova}), but still a
discrepancy of about a factor of $2$ in the best case remains.

Including a cosmological constant helps at large scales, but
decreases the velocities on small scales. 

If the observational bulk velocity data is indeed reliable (there
are some doubts about this\cite{Mark}), all global $O(N)$ models are 
ruled out.

\section{Conclusions}
We have developed a self contained formalism to determine CMB
anisotropies and other power spectra for models with causal scaling
seeds. We have applied it to global $O(N)$ models which contain global
monopoles and texture. Our main results can be summarized as follows:
\begin{itemize}
\item Global $O(N)$ models predict a flat spectrum
(Harrison-Zeldovich) of CMB anisotropies on large scales which is in
good agreement with the COBE results. Models with vanishing cosmological
constant and a large value of the Hubble parameter give $\si_8\sim
0.4$ to $0.5$ which is reasonable.
\item Independent of cosmological parameters, these models do not
exhibit pronounced acoustic peaks in the CMB power spectrum.
\item The dark matter power spectrum from  global $O(N)$ models with
$\Om_\La=0$ has reasonable amplitude but does not agree in its shape 
with the galaxy power spectrum, especially on very large 
scales $>20h^{-1}$Mpc.
\item Models with considerable cosmological constant agree relatively
well with the shape of the  galaxy power spectrum, but need very high 
 bias $b\sim 4 - 6$ even with respect to IRAS galaxies.
\item  The large scale bulk velocities are by a factor of about 3 to 5 
smaller than the value inferred from \cite{deke}.
\end{itemize}
In view of the still considerable errors in the CMB data (see
Fig.~\ref{fig24}), and the biasing problem for the dark matter power spectrum, 
we consider the last argument as the most convincing one to rule out  
global $O(N)$ models. Even if velocity data is still quite uncertain,
observations generally agree that bulk velocities on the scale of
$50h^{-1}$Mpc are substantially larger than the (50 -- 70)km/s 
obtained in texture models. 

However, all our constraints have been obtained assuming Gaussian
statistics. We know that global defect models are non-Gaussian, but
we have not investigated how severely this influences the above
conclusions. Such a study, which we plan for the future, requires 
detailed maps of fluctuations, the
resolution of which is always limited by computational resources. 
Generically we can just say that non-Gaussianity can only weaken the
above constraints.

Our results naturally lead to the question  whether all scaling seed models
are ruled out by present data.  The main problem of the $O(N)$
model is the missing power at intermediate scales, $\ell\sim 300 -
500$ or $R\sim (20 -100)h^{-1}\Mpc$. We have briefly investigated whether
this problem can be mitigated in a scaling seed model without vector
and  tensor perturbations. In this case, also scalar anisotropic
stresses are reduced by causality requirements (see Ref.~\cite{DK}), and the 
compensation mechanism mentioned in Section~II is effective. 
For simplicity, we analyze a model with purely scalar perturbations
and no anisotropic stresses at all, $f_\pi=0$. The seed function
$\Phi_s$ is taken from the texture model (numerical simulations) and
we set $\Psi_s=-\Phi_s$. The resulting CMB anisotropy spectrum is
shown in Fig.~\ref{fig19}, top panel. A smeared out acoustic peak with
an amplitude of about $2.2$ does indeed appear in this model. This
is mainly due to the fact that fluctuations on large scales are smaller in
this model, as is also evident from the higher value of 
$\ep=(2.2\pm 0.2)\times 10^{-5}$. But also here, the dark
matter density fluctuations and bulk velocities are substantially
lower than observed galaxy density fluctuations or the POTENT bulk
flows.

Clearly, this simple example is not sufficient and a more thorough 
analysis of generic scaling seed models is presently
under investigation. So far it is just clear that contributions from
vector and tensor perturbations are severely restricted.
\vspace{0.6cm}\\
{\large\bf Acknowledgment}\\
It is a pleasure to thank Andrea Bernasconi, Paolo de Bernardis,
 Roman Juszkiewicz, Mairi Sakellariadou, Paul Shellard, Andy Yates 
and Marc Davis for stimulating discussions. Our
Boltzmann code is a modification of a code worked out by the group
headed by Nicola Vittorio. We also
thank Michael Vogeley who provided us with the galaxy power spectra shown in our
figures.
The numerical simulations have been performed at the Swiss
super computing center CSCS. This work is partially supported by the
Swiss National Science Foundation.

\begin{onecolumn}

\begin{table}
\footnotesize
\begin{center}
\begin{tabular}{||c|c|c|c|c|c|c|c||}
\hline\hline
Experiment & Data point   &  $\Delta T^2 (\mu K)^2$ & 
$ + (\mu K)^2$ & 
$ - (\mu K)^2 $ &Sky Coverage 
& Reference& $\chi^2_j$ \\ 
\hline
COBE1 & 1& 25.2 & 183 & 25.2 & 0.65 & \protect\cite{tegma}& 125.29 \\
COBE2 & 2& 212 & 126 & 128 & 0.65 & \protect\cite{tegma}& 0.02 \\
COBE3 & 3& 256 & 96.5& 96.9& 0.65& \protect\cite{tegma}& 0.49 \\ 
COBE4 & 4& 105.5 & 48.3&48.2 & 0.65& \protect\cite{tegma}& 0.74\\
COBE5 & 5& 101.9 & 26.5& 26.4& 0.65& \protect\cite{tegma}& 0.1\\ 
COBE6 & 6& 63.4 & 19.11& 18.9& 0.65& \protect\cite{tegma}& 1.11\\ 
COBE7 & 7& 39.6 & 14.5 & 14.5& 0.65&\protect\cite{tegma} & 2.55\\ 
COBE8 & 8& 42.5 &12.7& 12.8& 0.65&\protect\cite{tegma} & 0.04\\ \hline
ARGO Hercules& 1& 360& 170& 140& 0.0024&\protect\cite{deb}& 0.001\\ \hline
MSAM93 & 2&4680&4200&2450&0.0007& \protect\cite{che}& 0.74\\ \hline
MSAM94 & 3& 4261 & 4091 & 2087  & 0.0007& \protect\cite{che2}&0.51\\
MSAM94 & 4& 1960 & 1352 &  858  & 0.0007& \protect\cite{che2} &0.01\\ \hline
MSAM95 & 5& 8698 & 6457 & 3406  & 0.0007& \protect\cite{che3} &1.47\\
MSAM95 & 6& 5177 & 3264 & 1864  & 0.0007& \protect\cite{che3} &0.30\\ \hline
MAX HR& 7& 2430& 1850&1020&0.0002&\protect\cite{tan}&0.001\\
MAX PH& 8& 5960& 5080&2190&0.0002&\protect\cite{tan}&0.41\\ 
MAX GUM& 9& 6580& 4450&2320&0.0002&\protect\cite{tan}&0.73\\ 
MAX ID& 10& 4960& 5690&2330&0.0002&\protect\cite{tan}&0.17 \\ 
MAX SH& 11& 5740& 6280&2900&0.0002&\protect\cite{tan}& 0.25\\ \hline
Tenerife & 12& 3975 & 2855 & 1807 & 0.0124& \protect\cite{gut} & 0.64\\ \hline
South Pole Q& 13& 480& 470& 160 & 0.005& \protect\cite{gun} &0.52 \\ 
South Pole K& 14& 2040& 2330& 790 & 0.005& \protect\cite{gun} & 0.01\\ \hline
Python& 15& 1940& 189& 490& 0.0006& \protect\cite{dra} & 0.37 \\ \hline
ARGO Aries& 16& 580& 150& 130& 0.0024& \protect\cite{mas}& 0.78\\ \hline
Saskatoon&17&1990&950&630&0.0037&\protect\cite{net}&0.79\\ 
Saskatoon&18&4490&1690 & 1360&0.0037&\protect\cite{net}&3.83\\
Saskatoon&19&6930&2770&2140&0.0037&\protect\cite{net}&4.60\\ 
Saskatoon&20&6980&3030&2310&0.0037&\protect\cite{net}&4.01\\ 
Saskatoon&21&4730&3380&3190&0.0037&\protect\cite{net}&1.32\\ \hline
CAT1 & 22& 934 & 403 & 232  & 0.0001&\protect\cite{bak} &1.36\\
CAT2 & 23& 577 & 416 & 238  & 0.0001&\protect\cite{bak} &0.62\\ \hline
\hline
\end{tabular}
\end{center}
\vspace{0.3cm}
\caption{The CMB anisotropy detections used in our analysis. The 3.,
4. and 5. column denote the value of the anisotropy and the upper and
lower 1-$\si$ errors respectively.
The references are: 
Tegmark and Hamilton 1997 \protect\cite{tegma}; de Bernardis {\it et
al.} 1994 \protect\cite{deb}; Cheng {\it et al.} 1994 \protect\cite{che}; 
Cheng {\it et al.} 1996
 \protect\cite{che2};  Cheng {\it et al.} 1997 \protect\cite{che3}; 
 Tanaka {\it et al.} 1996 \protect\cite{tan}; Gutierrez {\it et al.} 1997 
\protect\cite{gut}; Gundersen {\it et al.} 1993 \protect\cite{gun}; 
 Dragovan {\it et al.} 1993 \protect\cite{dra}; Masi et al 1996
\protect\cite{mas}; Netterfield {\it et al.} 1996 \protect\cite{net};
 Scott {\it et al.} 1997 \protect\cite{bak}.
 \label{tabdata}}
\end{table}

\begin{table}
\begin{center}
\begin{tabular}{||c|c|c|c|c|c||}
\hline
Catalog & $h$ & $\Omega_{\Lambda}$ & Best fit bias $b$& $\chi^2$&Data points\\
\hline
 CfA2-SSRS2 101 Mpc& 0.5 & 0.0 & 3.4 & 29 & 24\\
 CfA2-SSRS2 101 Mpc& 0.8 & 0.0 & 2.0 & 40 & 24\\
 CfA2-SSRS2 101 Mpc& 1.0 & 0.0 & 1.9 & 44 & 24\\
 CfA2-SSRS2 101 Mpc& 0.5 & 0.4 & 3.9 & 17 & 24\\
 CfA2-SSRS2 101 Mpc& 0.5 & 0.8 & 9.5 &  4 & 24\\
\hline
 CfA2-SSRS2 130 Mpc& 0.5 & 0.0 & 5.3 &  8 & 19\\
 CfA2-SSRS2 130 Mpc& 0.8 & 0.0 & 3.4 & 15 & 19\\
 CfA2-SSRS2 130 Mpc& 1.0 & 0.0 & 3.4 & 16 & 19\\
 CfA2-SSRS2 130 Mpc& 0.5 & 0.4 & 5.6 &  5 & 19\\
 CfA2-SSRS2 130 Mpc& 0.5 & 0.8 & 11.1 &  4 & 19\\
\hline
 LCRS  &0.5 & 0.0 & 3.0 &  71 & 19\\
 LCRS  & 0.8 & 0.0 & 1.8 & 96 & 19\\
 LCRS  & 1.0 & 0.0 & 1.6 & 108 & 19\\
 LCRS  & 0.5 & 0.4 & 3.7 &  33 & 19\\
 LCRS  & 0.5 & 0.8 & 8.7 &  40 & 19\\
\hline
 IRAS  &0.5 & 0.0 & 2.3 &  102 & 11\\
 IRAS  & 0.8 & 0.0 & 1.3 & 131 & 11\\
 IRAS  & 1.0 & 0.0 & 1.3 & 140 & 11\\
 IRAS  & 0.5 & 0.4 & 2.8 &  70 & 11\\
 IRAS  & 0.5 & 0.8 & 6.3 &   9 & 11\\
\hline
 IRAS 1.2 Jy & 0.5 & 0.0 & 4.2 &  56 & 29\\
 IRAS 1.2 Jy & 0.8 & 0.0 & 2.9 & 92 & 29\\
 IRAS 1.2 Jy & 1.0 & 0.0 & 2.9 & 99 & 29\\
 IRAS 1.2 Jy & 0.5 & 0.4 & 4.3 &  39 & 29\\
 IRAS 1.2 Jy & 0.5 & 0.8 & 6.7 &   28 & 29\\ 
\hline
 APM & 0.5 & 0.0 & 3.3 & 1350 & 29\\
 APM & 0.8 & 0.0 & 1.8 & 1500 & 29\\
 APM & 1.0 & 0.0 & 1.7 & 1466 & 29\\
 APM & 0.5 & 0.4 & 3.5 & 1461 & 29\\
 APM & 0.5 & 0.8 & 6.2 & 1500 & 29\\
\hline
 QDOT & 0.5 & 0.0 & 4.3 & 32 & 19\\
 QDOT & 0.8 & 0.0 & 2.9 & 44 & 19\\
 QDOT & 1.0 & 0.0 & 2.9 & 46 & 19\\
 QDOT & 0.5 & 0.4 & 4.3 & 25 & 19\\
 QDOT & 0.5 & 0.8 & 7.3 & 14 & 19\\
\hline
\end{tabular}
\end{center}
\vspace{0.3cm}
\caption{\label{tabpow}
Analysis of the matter power spectrum. In the first column the
catalog is indicated. Cols.~2 and 3 specify the model parameters. In
cols.~4 and 5 we give the bias parameter inferred by $\chi^2$
minimization as well as the value of $\chi^2$. Col.~6 shows the number
of 'independent' data points assumed in the analysis.}
\end{table}

\begin{table}
\begin{center}
\begin{tabular}{||c|c|c|c|c|c||}
\hline
R & $\sigma_v$ (R) & $\Delta_v$ & $h=0.5$& $h=1.0$& $\Omega_{\La}=0.8$\\
\hline
 10 & 494 & 170 & 145 & 205&  86 \\
 20 & 475 & 160 & 100 & 134&  78 \\
 30 & 413 & 150 &  80 &  98&  70 \\
 40 & 369 & 150 &  67 &  78&  65 \\
 50 & 325 & 140 &  57 &  65&  61 \\
 60 & 300 & 140 &  50 &  56&  57 \\
\hline
\end{tabular}
\end{center}
\vspace{0.3cm}
\caption{\label{tabevl}
Bulk velocities: Observational data from \protect\cite{deke} and theoretical
predictions. $\De_v$ estimates the observational uncertainty.
The uncertainties on the theoretical predictions are
around $\sim 30 \%$. The models $\Om_\La=0$ with $h=0.5$ and $h=1$ as well
as $\Om_\La=0.8,~h=0.5$ are investigated.}
\end{table}

\end{onecolumn}

\begin{twocolumn}

\appendix

\section{Complete definitions of gauge-invariant perturbation variables}

In this Appendix we give precise definitions of all the
gauge-invariant perturbation variables used in this paper. These
definitions, their geometrical interpretation and a short derivation
of the perturbation equations can be found in \cite{Review,DS}.
We restrict the analysis to the spatially flat case, $K=0$.
We define the perturbed metric by
\be g = \bar{g} +a^2h ~,\ee
where $\bar{g}$ denotes the standard Friedmann background, $a$ is the
scale factor and $h$ denotes the metric perturbation.

\subsection{Scalar perturbations}
Scalar perturbations of the metric are of the form
\bea h^{(S)} &=& -2A(dt)^2 + 2iBk_jdtdx^j + 2(H_L +{1\over 3}H_T)
\de_{ij}dx^idx^j \nonumber \\
 &&         -2k^{-2}H_Tk_ik_jdx^idx^j  ~ .  
\label{2h} \eea
Computing the perturbation of the Ricci curvature scalar and the shear
of the equal time slices, we obtain
\be 
   \de\bm{R} = 4 a^{-2}k^2{\cal R} ~\mbox{ , with }~~~ {\cal R} =
     H_L + {1\over 3} H_T  \label{2R} \; ,
\ee
\bea 
 K^{\mathrm (aniso)} &=& a\si({k_ik^j\over k^2} - {1\over
3}\de_i^j)dx^i\otimes\dd_j~,   \\
\mbox{with  } && \nonumber \\
    \si &=& k^{-1}\dot{H}_T - B  ~ . \label{2sigma} 
\eea
The Bardeen potentials are the combinations
\bea
 \Phi &=& {\cal R} - (\dot{a}/a)k^{-1}\si   \label{2Phi} \\
 \Psi &=&  A  - k^{-1}[(\dot{a}/a)\si -\dot{\si}] \; . \label{2Psi}
\eea
They are invariant under infinitesimal coordinate transformations
(gauge transformations).

To define perturbations of the most general energy momentum tensor, we
introduce the energy density $\rho$ and the energy flux $u$ as the
time-like eigenvalue and normalized eigenvector of $T^\mu_\nu$,
\[ T_{\mu}^{\;\;\nu}u^{\mu} = -\rho u^{\nu} \;\;,\;\; u^2 = -1 \; .\]
We then define the perturbations in the energy density and energy
velocity field by
\be \rho = \overline{\rho}(1+\de)  \;\;, \label{2de} \ee
\be u    = u^0\dd_t +u^i\dd_i ~;     \label{2v} \ee
$u^0$ is fixed by the normalization condition, $u^0=a^{-1}(1-A)$.
In the 3--space orthogonal to $u$ we define the stress tensor by
\be \tau^{\mu\nu} \equiv P^{\mu}_{~~\al}P^{\nu}_{~~\beta}T^{\al\beta}~,\ee
where $P= u\otimes u + g$ is the projection onto the sub--space of
$T{\cal M}$ normal to $u$. It is
\[ \tau^0_0 = \tau^0_i =\tau^i_0 = 0 ~ .\]
The perturbations of pressure and anisotropic stresses can be parameterized by
\be \tau_i^{\;j} = \bar{p}[(1+\pi_L  )\de_i^{\;j} + \pi_i^{\;\;j} ]
  ~~\mbox{ , with }~ \pi^i_i = 0 \;. \label{2pi} \ee
For scalar perturbations we set
\bean   
 u^0 &=& (1-A) \;,\; {u^{(S)j}\over u^0} = -i{k^j\over k}v \\
 \mbox{ and }  && \\
\pi^{(S)i}_{~j} &=&
   (-k^{-2}k^{i}k_j+{1\over 3}\de^i_{~j})\Pi~.
\eean
Studying the behavior of these variables under gauge transformations,
 one finds that the anisotropic stress potential $\Pi$ is gauge
 invariant. A gauge invariant velocity variable is the shear of the
   velocity field,
\be
\si^{(Sm)}_{ij} =(k^{-2}k_ik_j-{1\over 3}\de_{ij}a^3V  
	\mbox{ , ~with }~~ V =v-k^{-1}\dot{H}_T .
\ee

There are several different useful choices of gauge invariant density
perturbation variables,
\bea
  D_s &=& \de +3(1+w)(\dot{a}/a)k^{-1}\si \\
  D_g &=& \de +3(1+w){\cal R} = D_s +3(1+w)\Phi\\
  D   &=& D_s +3(1+w)(\dot{a}/a)k^{-1}V  \; .  \eea
In this work we mainly use $D_g$. Here $w=p/\rho$ denotes the enthalpy.
Clearly, these matter variables can be defined for each matter
component separately. For ideal fluids like CDM or the baryon photon
fluid long before decoupling, anisotropic stresses vanish and
$\pi_L=(c_s^2/w)\de$, where $c_s$ is the adiabatic sound speed.

Also scalar perturbations of the photon brightness, $\iota^{(S)}$ are not gauge
invariant. It has been show\cite{Review} that the combination 
\be
\MM^{(S)} =\io^{(S)} +4{\cal R} +4ik^{-1}n^jk_j\si  \label{Mio}
\ee
is gauge invariant. This is the variable which we use here. In other
work\cite{HS} the gauge invariant variable $\Th \equiv \MM+\Phi$ has
been used.  Since $\Phi$ is independent of the photon direction $\bn$
this difference in the definition shows up only in the monopole,
$C_0$.  But clearly,
as can be seen from Eq.~(\ref{Mio}), also the dipole, $C_1$, is gauge
dependent.

The brightness perturbation of the neutrinos is defined the same way
and will not be repeated here.
  
\subsection{Vector perturbations}
Vector perturbations of the metric are of the form
\be
 h^{(V)} = 2B_jdx^jdt +   ik^{-1}(k_lH_j+k_jH_l)dx^ldx^j
  \label{2hv}~,
\ee
where $\bf B$ and $\bf H$ are transverse vector fields. The simplest
 gauge invariant variable describing the two vectorial degrees of
freedom of metric perturbations is $\bm{\Si}$,
\be  \Si_j =k^{-1}\dot{H}_j-B_j ~.  \ee
Vectorial anisotropic stresses are gauge invariant. They are of the form
\be  \pi_{lj}^{(V)} = ik^{-1}(k_j\Pi_l+k_l\Pi_j)   ~.\ee
The vector degrees of freedom of  the velocity field are cast in the
vorticity
\bea
u_{l:j}-u_{j;l} &=& ia(k_j\om_l-k_l\om_j)  \\
\mbox{with} &&  \om_j=v_j-B_j ~.
\eea
Vector perturbations of the photon brightness are gauge-invariant.
To maintain a consistent notation, we denote them by $\MM^{(V)}$.
\subsection{Tensor perturbations}
We define tensor perturbations of the metric by
\be
 h^{(T)} = 2H_{ij}dx^idx^j
  \label{ht}~,
\ee
where $H_{ij}$ is a traceless transverse tensor field.

The only tensor perturbations of the energy momentum tensor are
anisotropic stresses,
\be  \pi_{lj}^{(T)} = \Pi_{lj}  ~.\ee
Tensor perturbations of the photon brightness are denoted
$\MM^{(T)}$.

Clearly, all tensor perturbations are gauge-invariant (there are no
tensor type gauge transformations).

\section{The CMB anisotropy power spectrum}

Here we derive in some detail Eqs.~(\ref{ClS}), (\ref{ClV}) and
(\ref{ClT}). 

CMB anisotropies are conveniently expanded in spherical
harmonics: $\delta T (\bn)/T_0 = \sum_{lm} a_{lm} Y_m^l
 (\bn)$.
The coefficients $a_{lm}$ are random variables with zero mean and
rotationally invariant
variances, $C_\ell \equiv \langle \mid a_{lm} \mid ^2 \rangle$.
The mean (over the ensemble) correlation function of the anisotropy 
pattern has the
standard expression:
\be
 \left\langle {\delta T\over T_0}(\bn_1){\delta T\over
T_0}(\bn_2)\right\rangle = {1\over 4\pi} \sum_\ell (2\ell+1) {C_\ell} P_\ell
(\cos\theta)
\ee
 where $\cos\theta = \bn_1 \cdot \bn_2$.
To find Eq.~(\ref{ClS}) we use the Fourier transform normalization
\be
  \hat{f}(\bk) = {1\over V}\int f({\bf x})\exp(i\bk\cd{\bf x})d^3x~,
\ee
with some normalization volume $V$. Assuming that ensemble average can
be replaced by volume average then implies
\bea
\lefteqn{\left\langle {\delta T\over T_0}(\bn_1)
 {\delta T\over T_0}(\bn_2)\right\rangle ~ = ~
{1\over V}\int d^3x  {\delta T\over T_0}({\bf x},\bn_1)
	{\delta T\over T_0}({\bf x}, \bn_2)} \nonumber \\
 &&= ~ {1\over (2\pi)^3}\int d^3k {\delta T\over T_0}(\bk,\bn_1)
	{\delta T\over T_0}(\bk, \bn_2)~.
\eea
Inserting our ansatz (\ref{expMM}) for  ${\delta T\over T_0} ={1\over 4}\MM$,
and using the addition theorem for spherical harmonics, we have
\bea
\lefteqn{ \left\langle {\delta T\over T_0}(\bn_1){\delta T\over T_0}(\bn_2)\right\rangle
=} \nonumber \\
&& {1\over 8\pi}\sum_{\ell,\ell',m,m'}(-1)^{(\ell-\ell')}Y_{\ell m}(n_1)
    Y^*_{\ell' m'}(n_2)\times  \nonumber \\
 &&  \int k^2dkd\Om_{\hat{\bk}}Y_{\ell m}^*(\hat{\bk})Y_{\ell' m'}(\hat{\bk})
  \langle\si_\ell\si^*_{\ell'}\rangle(k)  ~ = ~\nonumber \\
&&  ={1\over 32\pi^2}\sum_\ell(2\ell+1)P_\ell(\bn_1\cd\bn_2) \int
	k^2dk \langle\si_\ell\si^*_\ell\rangle(k)~, 
\label{ClSa}\eea
from which we can read of Eq.~(\ref{ClS}).

For vector and tensor fluctuations, the ansatz
(\ref{M12}) and (\ref{M+x}) must be taken into account.
With the same manipulations as above the correlation function of 
CMB anisotropies induced by vector modes reads:

\bea \lefteqn{\left\langle{\delta T\over T_0}(\bn_1){\delta T\over T_0}(\bn_2) 
	\right\rangle =} \nonumber \\
 && {1 \over {128 \pi^3}}
 \int d^3k\sum_{\ell 1 \ell 2}  \Pi_{\ell 1, \ell 2}
(\bk,{\bn'}_1,{\bn'}_2)
 P_{\ell 1} ({\mu'}_1) P_{\ell 2}({\mu'}_2)
\eea
where the primes indicate that the quantity is considered in the
reference system where $ \bk$ is parallel to the $z$ axis and
(\cite{koso},\cite{melvit})
\bea \Pi_{\ell 1,\ell 2} &=&
(-i)^{(\ell1-\ell2)}(2\ell_1+1)(2\ell_2+1)
\nonumber \\  &&
 \sqrt{(1 - (\mu'_2)^2)(1 - (\mu'_2)^2)} \times\nonumber \\  &&
[\langle \sigma^{(V)}_{1,\ell 1}\sigma_{1,\ell 2}^{(V) *}\rangle
 \cos (\phi'_1) \cos (\phi'_2) + \nonumber \\  &&
+ \langle \sigma_{2, \ell 1}^{(V)}\sigma_{2, \ell 2}^{(V) *}\rangle
 \sin (\phi'_1) 
\sin (\phi'_2)]~.
\eea

Assuming statistical isotropy which implies 
\[ \langle|\sigma_{1,\ell}^{(V)}|^2\rangle=
\langle|\sigma_{2,\ell}^{(V)}|^2\rangle ~~~\mbox{ and }~~~
  \langle\sigma_{1,\ell}^{(V)}\sigma_{2,\ell}^{*(V)}\rangle = 0 ~,\]
we obtain 
\bea
\lefteqn{\left\langle {{\delta T(\bn_1)}\over{T_0}}{{\delta T(\bn_2)}\over{T_0}} 
\right\rangle =} \nonumber \\
&& {1 \over {128 \pi^3}}\sum_{\ell 1 \ell 2}
( 2\ell_1 +1)(2 \ell_2 + 1) (-i)^{(\ell 1 - \ell 2)}\times\\
&& \int\Upsilon ({\bn'}_1, {\bn'}_2) 
\langle\sigma_{1,\ell 1}^{(V)}\sigma_{1,\ell 2}^{(V) *}\rangle
P_{\ell 1}({\mu'}_1)P_{\ell 2}({\mu'}_2) d^3 k~,
\eea
where
\be
\Upsilon ={{\bn_1} \cdot {\bn_2} - \mu'_1 \mu'_2}~.
\ee 
 Using the recursion formula for Legendre polynomials and the addition
  theorem for spherical harmonics, we find after some manipulations
\be
C_\ell^{(V)} = {\ell(\ell+1) \over {8 \pi}}  \int k^2dk{{\langle|
\si^{(V)}_{1,\ell+1}(t_0,k)+\si^{(V)}_{1,\ell-1}
(t_0,k)|^2\rangle}\over {(2 \ell+1)^2}} ~.
\label{ClVa}\ee

For the correlation function of the CMB anisotropies from tensor
modes our ansatz (\ref{M+x}) gives
\bea
 \lefteqn{{\left\langle {\delta T\over T_0}(\bn_1){\delta T\over T_0}(\bn_2)
\right\rangle} =} \nonumber \\
&& {1 \over {128 \pi^3}}
 \int {\sum_{\ell 1 \ell 2}  \Pi_{\ell 1, \ell 2}
(k,{\bn'}_1,{\bn'}_2)
 P_{\ell 1} ({\mu'}_1) P_{\ell 2}({\mu'}_2)
 d^3 k }
\eea
with
\bea 
\Pi_{\ell 1,\ell 2} &= & (-i)^{\ell1-\ell2}(2\ell1+1)(2\ell2+1)
 (1 - (\mu'_1)^2) \times\nonumber \\  &&
(1 - (\mu'_2)^2)[\langle\sigma^{(T)}_{\times,\ell 1}
 \sigma_{\times,\ell 2}^{(T) *}\rangle\cos (2 \phi'_1) \cos (2 \phi'_2)
+ \nonumber \\  && 
 + \langle\sigma_{+, \ell 1}^{(T)} \sigma_{+, \ell 2}^{(T) *}\rangle
 \sin (2 \phi'_1) 
\sin (2 \phi'_2)] .
\eea
Here, statistical isotropy leads to
\bea
\lefteqn{\left\langle {\delta T(\bn_1)\over T_0}{\delta T(\bn_2)\over T_0} 
\right\rangle =} \nonumber \\   &&
{1 \over 128 \pi^3}\sum_{\ell1\ell2}(2\ell_1+1)(2\ell_2+1)
	(-i)^{(\ell1 - \ell2)}\times  \nonumber \\ 
&& \int\Upsilon (\bn'_1, \bn'_2) 
\langle\sigma_{+,\ell 1}^{(T)}\sigma_{+,\ell 2}^{(T) *}\rangle 
P_{\ell 1}(\mu'_1)P_{\ell 2}(\mu'_2) d^3 k
\eea
where
\be
\Upsilon =[2 ({\bn_1} \cdot {\bn_2} - \mu'_1 \mu'_2)^2
 -  (1-(\mu'_1)^2)(1-(\mu'_2)^2)]~.
\ee 
With straightforward but somewhat cumbersome manipulations, applying
the recursion formula for Legendre polynomials and the addition
theorem for spherical harmonics, we then obtain
\be {C^{(T)}_\ell}= {1\over 8\pi} 
{(\ell+2)!\over(\ell-2)!}\int_0^\infty
{\mid{\Sigma^{(T)}_{\ell}(k)}\mid^2 \over(2\ell+1)^2}k^2dk~,
\label{ClTa}\ee
with
\be
\Si^{(T)}_{\ell}={{\si^{(T)}_{{\epsilon},\ell-2}} \over {2 \ell -1}}
-{{2{(2 \ell +1)\si^{(T)}_{{\epsilon},\ell}}} \over {(2 \ell -1)
(2 \ell +3)}}
+{{\si^{(T)}_{{\epsilon},\ell+2}} \over {{2 \ell +3}}}~.
\ee
The formulas (\ref{ClSa}),(\ref{ClVa}) and (\ref{ClTa}) are used to
determine the CMB anisotropy spectrum.

\end{twocolumn}


\begin{thebibliography}{99}
\bibitem{PLMAP}See the web sites:\\
	{\tt http://astro.estec.esa.nl/SA-general \\
	\qquad /Projects/Planck/} \\
	and {\tt http://map.gsfc.nasa.gov/}
\bibitem{infla}W. Hu, N. Sugiyama and J. Silk, { Nature} 
	{\bf 386}, 37 (1995).
\bibitem{observ}R. Durrer, M. Kunz, C. Lineweaver and M. Sakellariadou, {
        Phys. Rev. Lett.} {\bf 79}, 5198 (1997). 
\bibitem{DK}R. Durrer and M. Kunz, { Phys, Rev. D} {\bf 57}, R3199 (1998).
\bibitem{PST}U. Pen, U. Seljak and N. Turok,  { Phys. Rev. Lett.}  
	{\bf 79}, 1611	(1997).
\bibitem{Avel}P.P. Avelino, E.P.S. Shellard, J.H.P. Wu and B. Allen,
	Phys. Rev. Lett., in print (archived under {\tt 
	astro-ph/9712008}) (1998).
\bibitem{ZD}R. Durrer and Z. Zhou, { Phys. Rev. D} {\bf 53}, 5394 (1996).
\bibitem{D90}R. Durrer, { Phys. Rev. D} {\bf 42}, 2533 (1990).
\bibitem{Review}R. Durrer, { Fund. of Cosmic Physics} {\bf 15}, 209
	(1994).
\bibitem{StW}J.M. Stewart and M. Walker,
	{ Proc. R. Soc. London} {\bf A341}, 49  (1974).
\bibitem{Ba}J. Bardeen, { Phys. Rev. D}
        {\bf 22}, 1882 (1980).
\bibitem{KS}H. Kodama and  M. Sasaki, { Prog. Theor. Phys. Suppl.} 
        {\bf 78}, 1 (1984).
\bibitem{DS}R. Durrer and M. Sakellariadou,  {
        Phys. Rev. D} {\bf 56}, 4480 (1997).
\bibitem{MC}C. Cheung and J. Magueijo,  { Phys. Rev. D} {\bf 56},
	1982 (1997).
\bibitem{UDT}J.P. Uzan, N. Deruelle and N. Turok, { Phys. Rev. D} {\bf 57},
	7192 (1998).
\bibitem{melvit}
	A. Melchiorri and N. Vittorio, {\em Polarization of the 
	microwave background:
	Theoretical framework}, in: Proceedings of the NATO Advanced 
	Study Institute 1996 on the "Cosmic Background Radiation", 
	Strasbourg, archived under  {\tt astro-ph 9610029}, (1996)
\bibitem{ABR}A. Albrecht, R. Battye and J. Robinson, {
	Phys. Rev. Lett.} {\bf 79}, 4736 (1997).
\bibitem{ACFM}A. Albrecht,  D. Coulson, P.G. Ferreira and
	J. Magueijo, { Phys. Rev. Lett.} {\bf 76}, 1413 (1996).
\bibitem{Aetal}B. Allen {\it et al.},  { Phys. Rev. Lett.} {\bf 79}, 2624
	(1997).
\bibitem{KD}M. Kunz and R. Durrer, { Phys. Rev. D} {\bf 55}, R4516
	(1997).
\bibitem{CHM}C. Contaldi, M. Hindmarsh and J. Magueijo, preprint
	archived under {\tt astro-ph/9808201} (1998).
\bibitem{TSlN}N. Turok and D. Spergel, Phys. Rev. Lett {\bf 66}, 3093 (1991).  
\bibitem{Ma}J. Magueijo, A. Albrecht, P.G. Ferreira and D. Coulson,
	{ Phys. Rev. D} {\bf 54}, 3727 (1996).
\bibitem{DGS}R. Durrer, A. Gangui and M. Sakellariadou, {
	Phys. Rev. Lett.} {\bf 76}, 579 (1996).
\bibitem{TC} R. Crittenden and N. Turok, Phys. Rev. Lett. {\bf 75},
	2642 (1995).
\bibitem{Paul} P. Shellard, private communication (1998).
\bibitem{debe}P. de Bernardis, A. Balbi, G. De Gasperis, 
	A. Melchiorri and N. Vittorio, 
	{ Astrophys. J. }{\bf 480}, 1 (1997).
\bibitem{maber}C.P. Ma and E. Bertschinger, { Astrophys. J.}
	 {\bf 455}, 7 (1995).
\bibitem{jones}B.J. Jones and R. Wyse, {Astron. \& Astrophys.} {\bf 149},  
	144  (1985).
\bibitem{peeyu}
	P.J.E. Peebles and J.T. Yu,{ Astrophys. J.} {\bf 162}, 815 (1970)
\bibitem{CMBfast}U. Seljak and M. Zaldarriaga, { Astrophys. J.}
	{\bf 496}, 437 (1996).
\bibitem{PST2}U. Pen, D. Spergel and N. Turok, { Phys Rev.} D {\bf 49},
	692 (1994).
\bibitem{aberna}W. P. Petersen and A. Bernasconi,
	{ CSCS Technical Report} TR-97-06 (1997).
\bibitem{JB} J. Borrill, { private communication}.
\bibitem{TS} N. Turok and D. Spergel, Phys. Rev. Lett. {\bf 64}, 2736
	(1990).
\bibitem{xiao}X. Luo,{ Astrophys. J.} {\bf 439},
	 517L (1995) .
\bibitem{ACSSV} B. Allen {{\it et al.}},  Phys. Rev. Lett 
	{\bf 77}, 3061 (1996).
\bibitem{tegma} M. Tegmark and A. Hamilton, archived under 
	{\tt astro-ph/9702019} (1997)
\bibitem{deb} P. de Bernardis {\it et al.}, { Astrophys. J.} {\bf 422}, 33L (1994).
\bibitem{che} E.S. Cheng {\it et al.}, { Astrophys. J.} {\bf 422}, 40L (1994).
\bibitem{che2} E.S. Cheng {\it et al.}, { Astrophys. J.} {\bf 456}, 71L (1996).
\bibitem{che3} E.S. Cheng {\it et al.}, \apj {\bf 488}, 59L (1997).
\bibitem{tan} S.T. Tanaka {\it et al.}, { Astrophys. J.} {\bf 468}, 81L (1996). 
\bibitem{gut} C. M. Gutierrez {\it et al.}, { Astrophys. J.} {\bf 480}, 83L (1997).
\bibitem{gun} J.O. Gundersen
	 {\it et al.}, \apj {\bf 413}, 1L (1993).
\bibitem{dra} M. Dragovan
	{\it et al.}, \apj {\bf 427}, 67L (1993). 
\bibitem{mas} S. Masi {\it et al.}, \apj {\bf 463}, 47L (1996).
\bibitem{net} B. Netterfield {\it et al.},  
	\apj {\bf 474}, 47 (1997).
\bibitem{bak} P.F. Scott at al., \apj {\bf 461}, 1L (1996).
\bibitem{eke}V.R. Eke, S. Cole and C.S. Frenk { M.N.R.A.S.} {\bf 282},
	 263E (1996). 
\bibitem{lid}A.R. Liddle, D. Lyth, H. Robeerts, P. Viana,
 	{ M.N.R.A.S.} {\bf 278}, 644 (1996). 
\bibitem{free} W. Freedman, J.R. Mould, R.C. Kennicut and B.F. Madore,
	archived under {\tt astro-ph/9801080} (1998).
\bibitem{fish} K.B. Fisher, M. Davis, M.A. Strauss, 
	A. Yahil and J.P. Huchra,{ M.N.R.A.S.}  {\bf 266}, 50 (1994).
\bibitem{tadr} H. Tadros and  G. Efstathiou, { M.N.R.A.S.},
	{\bf L45}, 276 (1995).
\bibitem{daco} L.N. Da Costa, M.S. Vogeley, M.J. Geller, J.P. Huchra
	and C. Park, { Astrophys. J.} {\bf 437}, 1L (1994).
\bibitem{lin} H. Lin {\it et al.}, {AAS}, {\bf 185}, 5608L (1994).
\bibitem{baugh} C.M. Baugh and  G. Efstathiou, {M.N.R.A.S.}, {\bf 265},
	145 (1993).
\bibitem{kais}N. Kaiser, Astrophys. J. {\bf 284},
 	9L (1984). 


\bibitem{kais2}N. Kaiser, G. Efstathiou, R. Ellis,
	C.S. Frenk, A. Lawrence, M. Rowan-Robinson and 
	W. Saunders, { M.N.R.A.S.}  {\bf 252}, 1 (1991). 
\bibitem{will}J.A. Willick, S. Courteau, S.M. Faber,
	D. Burstein, A. Dekel and T. Kolatt, 
	{ Astrophys. J.}  {\bf 457}, 460 (1996). 
\bibitem{pea97}J.A. Peacock, { M.N.R.A.S.}  {\bf 284},
	 885P (1997). 
 
\bibitem{shai} E.J. Shaia, P.J.E. Peebles and
	R.B. Tully, { Astrophys. J.}  {\bf 454}, 15 (1995). 
\bibitem{deke} A. Dekel, Ann. Rev. of Astron. and Astrophys. {\bf 32}, 371 (1994).


\bibitem{giova} R. Giovanelli {\it et al.}, preprint, archived under 
	{\tt astro-ph/98707274} (1998).
\bibitem{voegel} M.S. Vogeley, to appear in "Ringberg Workshop on
	Large-Scale Structure", ed. D. Hamilton, (Kluwer, Amsterdam),
	archived under {\tt astro-ph/9805160} (1998).


\bibitem{Mark}M. Davis, private communication (1998).
\bibitem{HS}W. Hu and N. Sugiyama, Phys. Rev. D {\bf 51}, 2599
	(1995), and references therein.
\bibitem{koso} A. Kosowsky, Annals of Physics {\bf 246}, 49 (1996).
\end{thebibliography}
\end{document}